\documentclass[12pt]{article}
\usepackage{amsmath, latexsym, amssymb, amscd}
\numberwithin{equation}{section}

\newcommand{\R}{\mathbb R}
\newcommand{\C}{\mathbb C}

\newcommand{\p}{\partial}

\newcommand{\Bold}[1]{{\boldsymbol{\mathit{#1}}}}

\newcommand{\M}{\mathbb{M}}

\newtheorem{proposition}{Proposition}[section]
\newtheorem{definition}{Definition}[section]

\newtheorem{remark}{Remark}[section]

\begin{document}

\title{Functional Integral Approach to $C^*$-algebraic Quantum Mechanics II:\\ Symplectic Quantum Mechanics}
\author{J. LaChapelle}

\maketitle

\begin{abstract}
We propose $Sp\,(8,\R)$ and its Langlands dual $SO(9,\R)$ as dynamical groups
for closed quantum systems. Restricting here to the non-compact group $Sp\,(8,\R)$, the quantum theory is
constructed and investigated. The functional Mellin transform plays
a prominent role in defining the quantum theory. It provides a
bridge between the quantum algebra of observables and the algebra of
operators on Hilbert spaces furnishing  unitary representations
that are induced from a distinguished parabolic subgroup of $Sp\,(8,\C)$.  As well, the parabolic subgroup renders a fiber bundle construction that models what can be described as a matrix quantum gauge theory. The formulation is strictly quantum mechanics: no \emph{a priori} space-time is assumed and the only geometrical input comes indirectly from the group manifold. But what appears on the surface to be a fairly simple-minded model turns out to have a capacious structure suggesting some compelling physical interpretations regarding space-time and fundamental interactions.
\end{abstract}

\section{Introduction}
\subsection{Motivation}
The program of quantization is most often approached from the bottom up. That is, physical considerations identify a classical phase space (whose points represent classical states) and suitable functions on that phase space. Then one attempts to promote: (i) the classical phase space to a suitable Hilbert space, and (ii) functions on the phase-space to suitable operators on the Hilbert space.

The obvious alternative is a top-down approach. Here the goal is to somehow construct, \emph{without appealing to a classical system}, a $C^\ast$-algebra and an associated Hilbert space. Of course, the key  is to find the correct formulation `up stairs', since generally it doesn't correspond one-to-one with the `down stairs' classical theory.

Our tack in this paper is a top-down approach that mingles both functional integral and algebraic elements. The idea is to model the $C^\ast$-algebra $\mathfrak{A}$ characterizing a quantum system by a $C^\ast$-algebra of integrable functionals\footnote{The functional $\mathrm{F}\in\mathbf{F}(G^\C)$ is said to be integrable if the associated functional integral is well-defined (to be explained later). We use the term ``functional'' to refer to an operator-valued function on some topological space. Strictly, functional refers to a scalar-valued function on some vector space. However, in the context where Banach-valued functions on topological spaces are integrands of functional integrals, we will continue to use the term imprecisely.} $\mathbf{F}(G^\C)\ni\mathrm{F}:G^\C\rightarrow L_B(\mathcal{H})$ where $G^\C$ is the complexification of a  topological group isomorphic to the group of units of $\mathfrak{A}$,  the Hilbert space $\mathcal{H}$ furnishes a direct sum of all relevant unitary representations of  $G^\C$, and $L_B(\mathcal{H})$ is the $C^\ast$-algebra of linear bounded operators on $\mathcal{H}$. Importantly, the two $C^\ast$-algebras $\mathbf{F}(G^\C)$ and $L_B(\mathcal{H})$ are related to each other via the functional Mellin transform (to be explained later).

This approach, which is related to quantization via crossed products, allows to view the main task of quantization as a choice of some topological group $G$. The topological group (along with the functional Mellin transform) simultaneously generates: (i) the Hilbert space of states, (ii) the $C^\ast$-algebra of integrable functionals, and (iii) the evolution dynamics.\cite{LA1}

It is important to emphasize that $G$ is generically not measurable. Insofar as a quantum system must be quantified through observation/measurement, $G$ must therefore be inferred from  \emph{locally compact} topological groups homomorphic to $G$. The idea is that observation/measurement of a quantum system corresponds to a homomorphism $\lambda:G\rightarrow G_\lambda$ with  $G_\lambda$ a locally compact topological group. In consequence, $G$ is indirectly identified with an entire family $G_\Lambda:=\{G_\lambda\,,\,\lambda\in\Lambda\}$ where the set $\Lambda$ characterizes all possible `localizations' $\lambda:G\rightarrow G_\lambda$ of a given system.\footnote{We are skipping a good deal of detail and explanation here that can be found in \cite{LA1} and \cite{LA4}.} These `localizations' embody the Born rule by reducing pertinent functional integrals to bona fide Haar integrals.

 Given its significant responsibility, one would expect a judicious choice of $G_\Lambda$ would lead to interesting and relevant physics if the functional approach is indeed valid. So the obvious next step in the program is to determine or otherwise guess $G_\Lambda$ and then study the resulting QM.

We first give a brief motivation regarding our guess for $G_\Lambda$. Begin with the simple idea that dynamical
interactions can be modeled by correlations among a set of
`constituents'.\footnote{By `constituents' we mean excitations of a
quantum system that are (quasi)stationary in time during an
evolution.} Consider a physical quantum system composed of $N$ such constituents. At a purely
formal level, the correlations representing
interactions among $N$ constituents leads to the identification of
$U(N)$ as a maximal organizing group. Under various \emph{dynamical}
circumstances,  certain correlations will dominate others, and this
will induce sub-organizations in $U(N)$ referred to as subduction. Evidently this subduction will continue to be driven by dynamics until some (quasi)equilibrium organizing group, which we will call the dynamical group, is achieved that describes the system correlations. A few moments reflection on the subduction of $U(N)$ into its
relevant subgroup chains that incorporate observed spin properties of space-time systems (along with some liberal hand-waving) leads to just two parent groups that seem
to be viable candidates for a quantum dynamical group
--- $Sp\,(8,\R)$ and $SO(9,\R)$ --- depending on odd or even permutation symmetry. These just happen to be Langlands dual.

Although it is likely that
$SO(2n+1,\R)$ and $Sp\,(2n,\R)$ are viable dynamical groups as well
for some interesting physical systems (especially for $n<4$), it is our contention that
$Sp\,(8,\R)$ and $SO(9,\R)$ are the most relevant for the majority of physical
observation at typical terrestrial energy densities. Our hand-waving motivation aside, we therefore \emph{postulate} that $Sp\,(8,\R)$ and $SO(9,\R)$ are dynamical groups that govern (some interesting) quantum systems.\footnote{Because of the close connection between $Sp\,(8,\R)$ and $SO(9,\R)$, it is tempting to bring them together under $OSp\,(9|8)$ to incorporate supersymmetry. However, our hunch is that
$OSp\,(9|8)$ is rather like a dual correlation/scattering description of an observable system, and the usual notion of supersymmetric states is not realized in nature.}

To keep the exposition manageable, we will restrict attention to $Sp\,(8,\R)$ in this paper and investigate only symplectic quantum mechanics (SQM).  Whether or not $Sp\,(8,\R)$ describes realistic quantum systems and reduces to realistic classical systems is of course paramount. We will present evidence to suggest that it does, but obviously the issue cannot be settled in a single paper.

\subsection{The symplectic case}

Symplectic symmetry is no stranger to classical mechanics, and the
suspicion that symplectic groups have importance as \emph{dynamical}
groups for quantum mechanics\footnote{The compact form of the symplectic group shows up in string theory too, but we are restricting attention to dynamical groups in the context of simple QM here.} has been around for a long time ---
for obvious reasons based on the correspondence principle and the
close relationship between the symplectic and Heisenberg groups (see
e.g. \cite{PR}). Many works rely on a discrete-series representation of $Sp(2n,\R)$. This has advantages and disadvantages: On one hand, it allows one to use familiar methods involving raising and lowering operators on discrete number-type states. On the other hand, in some applications the physical interpretation of the discrete states is not always manifest. In particular, it is not clear how to interpret the quantum numbers of the discrete states for $Sp(8,\R)$. Nevertheless, the idea of symplectic dynamical groups continues to attract
attention and produce reasonable successes; most notably perhaps in
nuclear physics (see \cite{RW} and the references therein).\footnote{$Sp(2n,\R)$ has been investigated in the context of quantum optics \cite{ADMS}, \cite{WU}; but as a symmetry of the canonical commutation relations not as a dynamical group.}

Having settled on $Sp(8,\R)$ as a dynamical group, the first task is to define the quantum theory. For this we utilize and assume familiarity with \cite{LA1}. Applying the quantization scheme proposed in \cite{LA1} invokes three key notions: (i) $Sp(8,\R)$ contains a distinguished parabolic subgroup that determines relevant induced unitary representations which are then used to construct the quantum Hilbert space. (ii) A $C^\ast$-algebra containing quantum observables is constructed via functional Mellin transforms. Together with the Hilbert space, this provides the kinematic backdrop of the quantum theory, and it allows concrete functional integral realizations of interesting quantum operators. (iii) Inner automorphisms of the symplectic group induce inner automorphisms of the $C^\ast$-algebra that yield system dynamics.

Granted the quantum theory is well-defined by this construction, we move on to interpret and explore some physical implications.

An important attribute of $Sp\,(8,\R)$ is it possesses a parabolic subgroup $P$ with  $\mathrm{dim}_\R(P)=26$ that contains the maximally compact subgroup $U(4)\subset P$. Since $Sp\,(8,\R)$ is supposed to be dynamical and the group elements that are not contained in $P$ mutually commute, we propose  that $P$ can be interpreted as a symmetry group responsible for internal forces associated with $U(4)$ and external forces associated with $P/U(4)$. (The terms `internal' and `external' are being used here in the conventional sense: There is strictly no internal/external dichotomy in SQM. But the two terms gain meaning within SQM pending an eventual space-time interpretation  of certain expectation values.)

 At the Lie algebra level, $\mathfrak{Sp}(8)$ comprises $36$ generators; ten of which mutually commute. Of the remaining $26$ that generate the parabolic subgroup $P$, sixteen span the algebra $\mathfrak{U}(4)$.\footnote{The standard notation for these Lie algebras would be $\mathfrak{sp}(8)$ and $\mathfrak{u}(8)$. We choose to maintain the capitalization of the associated Lie group to more clearly differentiate between the algebra and its elements for generic groups. For example, for a group we write $g\in G$ and for its algebra $\mathfrak{g}\in\mathfrak{G}$.} The remaining ten generators of $P$, in combination with the only two involutive inner automorphisms of the algebra, ultimately give rise to `observed geometry'. More precisely,  ground-state expectation values of certain operators coming from $P/U(4)$ characterize the geometry of a ten-dimensional manifold parametrized by the spectrum of the commuting generators associated with the homogenous space $X=Sp\,(8,\R)/P$ (under suitable conditions).

According to the manifold structure of $Sp\,(8,\R)$, there are five local domains where the radial group parameters (which parametrize the dual maximal torus) induce metric signatures $(0,4)$, $(1,3)$, $(2,2)$, $(3,1)$ and $(4,0)$ on the maximal torus, because the radial parameters can be real or imaginary.\cite{KM} The odd signatures give rise to what may be interpreted as $10$-d `observed space-time' for the configuration space of a certain sub-cotangent bundle (to be explained later). Furthermore, since the generators associated with $X$ and $P$ carry $U(4)$ charge and are dynamical, their spectra are likewise dynamical. Consequently, `observed space-time' is dynamical.

This notion of observed geometry is not unique to the symplectic group:  a similar statement can be made regarding the Heisenberg group in standard non-relativistic quantum mechanics. That is, VEVs of suitable combinations of operators can be interpreted as an `observed cotangent bundle' with a suitable choice of representation. The difference is that the observed geometry of $Sp(8,\R)$ leads to a $10$-d configuration space with a $U(4)$ internal symmetry. The ten dimensions aside, this difference is significant because it mixes the configuration parameters and $U(4)$ charges\footnote{We will consistently use the term `charge' to mean the eigenvalue of a charge operator.}. We will see later that this means the associated (possibly) time-dependent operators create/annihilate $U(4)$ charges at different points in $X$ precisely because of the non-abelian nature of $U(4)$. This can be interpreted as particle transmutation which of course does not happen in standard non-relativistic QM. It also means that matter (understood as the presence of $U(4)$ charge) and geometry are inseparable in SQM at the quantum algebra level.

The parabolic subgroup plays a second important role: it is the basis of induced representations which find a natural description in a fiber bundle framework. The induced representations ultimately form the quantum Hilbert space of states and the associated fiber bundle geometry gives a coherent state\footnote{The term coherent state is a bit imprecise, but we
will conform to standard usage. Strictly, for us a coherent state is
a model of a Hilbert \emph{state vector} (as opposed to a
projective-Hilbert \emph{state}) on a (sub)manifold of some Lie
group.} (CS) model of state vectors and operators. It is in the CS formulation where the physical interpretation of the theory begins to emerge: one can interpret/identify ground states, matter and gauge particles, and their associated fields; as well as exhibit explicit realizations of relevant operators.\footnote{Although indirectly related, the coherent state model of fields and operators are not the same fields and operators of a QFT.\cite{LA1}} With these objects, meaningful transition amplitudes can be constructed and interpreted.

Here again, the idea to use inducing representations is not unique to the symplectic group: the same approach is used for the non-compact Poincar\'{e} group. In the case of Poincar\'{e}, the little group and the mutually commuting momentum generators yield spin/helicity and momenta labels for state vectors along with a particle creation/annihilation interpretation; and they induce an `observed fiber bundle' structure with a \emph{momentum} base space and Lorentz structure group.\footnote{For Poincar\'{e}, one can then construct an `observed phase space' by attaching a static space-time manifold to the momentum base space of the `observed fiber bundle'. Recall that the boost generators contain all the dynamics induced by the energy operator: the $4$-momentum and angular-momentum generators are inert. Accordingly, the `observed phase space' is static (assuming the free Hamiltonian is derived from the momentum operators), and it would seem the dynamics associated with boost can be naturally interpreted as inertia.} The crucial difference brought by $Sp(8,\R)$ --- beyond the ten dimensions ---  is the mingling of `internal' and `external' symmetries resulting in a dynamical  bundle structure.

Besides providing physical interpretation via a distinguished parabolic structure, the symplectic group is supposed to govern system dynamics through inner automorphisms of its induced $C^\ast$-algebra. It turns out that the CS model of such dynamics in the Heisenberg picture is matrix quantum mechanics --- which is known to have a deceptively intricate structure. For example, it is known that in the adjoint representation the dynamics of the ten commuting operators of the Lie algebra approach a membrane theory for large systems. The remaining operators, which generate the parabolic subgroup, represent gauge degrees of freedom, and the adjoint representation provides a matrix gauge theory interpretation.

Because the dynamics is governed by inner automorphisms, evolution transforms the parabolic subgroup $P$ according to the adjoint action. In consequence, for non-trivial dynamics determined by some time-dependent $h(t)\in Sp(8,\R)$ a new, more relevant parabolic subgroup $\widetilde{P}=Ad(h(t_0))P$ can emerge at some time $t_0$ in the evolution epoch. Nevertheless, the induced representations associated with $\widetilde{P}$ are unitarily equivalent to those induced by $P$.\cite{LA1} However, the CS model of state vectors depends on $\widetilde{X}$. In particular this applies to CS ground states. Hence, a new non-trivial (relative to $P$) ground state may be associated with the final state of an evolution process by a suitable choice of representation.\footnote{The
notion of a CS model with a non-trivial ground state is similar in spirit
to an effective theory of quasi-particles. However there is an
important difference: all ground states (along with their associated CS) are
related via evolution through
$Sp\,(8,\R)$.} The catch is that one must somehow relate the
original ground state representation to the non-trivial ground state
representation in order to relate and interpret the physical
properties of the corresponding coherent states. Nevertheless, the
underlying framework is a quantum theory, it has a dynamically induced adjustable ground state, and it applies equally to all
resolution scales; micro-, meso-, and macroscopic systems --- assuming one is clever enough to identify the pertinent ground states.

Having proposed interpretations of various objects in the quantum theory, one would like to compare with corresponding classical objects. The associated classical dynamics is defined via the correspondence principle, and the classical Poisson bracket turns out to be the large-system (i.e. many-particle) limit of the Lie algebra-induced bracket on the $C^\ast$-algebra. The end result is classical Hamiltonian mechanics on a cotangent bundle with a $10$-d configuration space. Six of the dimensions are interpreted as directed-area degrees of freedom, and hence there is no incentive to compactify them. Indeed, they appear to have important physical significance: They naturally\footnote{Natural in the sense that the volume derives from a wedge product between expectations of \emph{two} commuting generators of the algebra associated with directed area; thus requiring just a two-component interaction. In fact the elementary geometric objects in $4$-d, i.e. $n$-cubes and their $n$-faces for $n\leq4$, can all be derived from products between expectations of just two suitable commuting generators of the group algebra. In other words, a formal wedge-product interaction between two quantum excitations associated with the commuting generators is enough to construct $4$-d geometric objects.

This is not the case if such objects are constructed from linear dimensions alone. In particular, constructing a volume element from commuting generators associated with only linear dimensions requires either four iterated wedge products or a formal wedge product between a generator (associated with a linear dimension) and an object (associated with the volume boundary) that does not belong to the group algebra.} define a volume in a $4$-d space, and they may represent vortex-like degrees of freedom.

One final point; $Sp(8,\R)$ has creation/annihilation representations; the infinite-dimensional, irreducible discrete-series representations that foster a fluctuation-type picture. So at the end of the day, one can throw out all the physical motivation and interpretation supplied by the parabolic decomposition and its concomitant coherent state model and simply refer everything to the discrete-series representation(s). Presumably, this would constitute a rigorous, geometry-free mathematical formulation of SQM --- physical interpretation notwithstanding.

\section{Quantization}
\subsection{Representations}
It is appropriate to begin with a
review of the structure of the symplectic algebra pointing out some
of its implications and then to construct unitary representations (URs) that are
particularly amenable to physical interpretation.

\subsubsection{Symplectic Lie algebra}\label{Lie algebra} $Sp\,(8,\R)$
is a rank-$4$ reductive Lie group of $\mathrm{dim}_\R(Sp\,(8,\R))=36$. The first
order of business is to examine the structure of the adjoint
representation to learn what type of dynamics it encodes.

Consider the triangular decomposition of its Lie algebra;
\begin{equation}
\mathfrak{Sp}(8)=:\mathfrak{G}=\mathfrak{G}_-\oplus\mathfrak{G}_0
\oplus\mathfrak{G}_+
\end{equation}
where
\begin{eqnarray}
&&\left[\mathfrak{G}_0,\mathfrak{G}_0\right]=0\notag\\
&&\left[\mathfrak{G}_+,\mathfrak{G}_-\right]\subseteq\mathfrak{G}_0\notag\\
&&\left[\mathfrak{G}_\pm,\mathfrak{G}_0\oplus\mathfrak{G}_\pm\right]
\subseteq\mathfrak{G}_\pm\;.
\end{eqnarray}
This decomposition is physically relevant, because it defines charges (quantum numbers) associated with the symmetry that
can be used to characterize states if the associated quantum system respects the symmetry. The subalgebra $\mathfrak{G}_0$ characterizes neutral states and $\mathfrak{G}_\pm$ characterizes charged states associated with `particles' and `anti-particles'. In the adjoint representation, there are four neutral states and $32$ charged states possessing various combinations of four types of charge.

To render these brackets more explicit, let $F_{f}$ denote a Fock
space of \emph{fermionic} excitations above some vacuum. Define creation and
annihilation operators acting on $F_{f}$ by
\begin{equation}
c_\alpha c^\dag_\beta+ c^\dag_\beta c_\alpha=\delta_{\alpha\beta}
\;\;\;;\;\;\; c^\dag_\alpha c^\dag_\beta+ c^\dag_\beta
c^\dag_\alpha=0\;\;\;;\;\;\; c_\alpha c_\beta+ c_\beta c_\alpha=0
\end{equation}
where $\alpha,\beta\in\{\pm1,\ldots,\pm4\}$. With these operators, a basis of
$\mathfrak{Sp}(8)$ can be realized as $\{c_{j,j},c_{j,-j},c_{-j,j},c_{j,j-i},c_{j-i,j}\}$ where
\begin{equation}
c_{j,i}:=\frac{1}{\sqrt{1+\delta_{j,-i}}}
\left(c^\dag_j c_i-(-1)^{j-i}c^\dag_{-i}
c_{-j}\right)
\end{equation}
with $j\in\{1/2,3/2,5/2,7/2\}$ and $i\in\{1,\ldots,2j-1\}$.\cite[\S\S5.9,5.20]{JQ}  The basis can be arranged as
\begin{eqnarray}\label{rearranged generators}
&&\mathfrak{h}_a:=c_{a,a}=:n_a-n_{-a}\notag\\
&&\mathfrak{e}_a:=c_{a,-a}=\sqrt{2}c_a^\dag c_{-a}\notag\\
&&\mathfrak{e}_{a,\pm b}:=c^\dag_a c_b\mp c^\dag_{-b}
c_{-a}
\end{eqnarray}
with $\mathfrak{e}_a^\dag=\mathfrak{e}_{-a}$ and $\mathfrak{e}^\dag_{b,\,a}=\mathfrak{e}_{a,b}$ where $a,b\in\{1,\cdots,4\}$ and $a< b$. This arrangement characterizes the Borel subgroup and its induced
coset space with associated subalgebras
$\mathfrak{G}_0\cong\mathrm{span}_\R\{\mathfrak{h}_a\}$,
$\mathfrak{G}_+\cong\mathrm{span}_\R\{\mathfrak{e}_a,\mathfrak{e}_{a,b}\}$, and
$\mathfrak{G}_-\cong\mathfrak{G}_+^\dag$.

The Borel decomposition can be used to build up a bosonic Fock space of states giving rise to  infinite-dimensional, irreducible positive discrete-series representations of $Sp(8,\R)$.\cite{DQ} The states clearly represent excitations due to the action of $\mathfrak{Sp}(8)$, but the interpretation of the excitations and their associated charges in relation to physical particles is not immediately clear. Instead, we will choose a parabolic decomposition motivated by the fact that $U(4)$ is the maximal compact subgroup and the observation that there are ten mutually
commuting generators contained in
$\mathfrak{G}_-\oplus\mathfrak{G}_+$.

Consider $\mathfrak{Sp}(8)
=\mathrm{span}\{\mathfrak{u}_{ab},\mathfrak{\mathfrak{e}}_{ab},\mathfrak{\mathfrak{e}}_{ab}^\dag\}$ where
\begin{eqnarray}
&&\{\mathfrak{u}_{ab}\}
:=\{\mathfrak{h}_a,\mathfrak{e}_{a,-b},\mathfrak{e}^\dag_{a,-b}\}
,\;\;\;\;\;1\leq a< b\leq 4\notag\\
&&\{\mathfrak{\mathfrak{e}}_{ab},\mathfrak{\mathfrak{e}}_{ab}^\dag\}:=\left\{(\mathfrak{e}_{a},\mathfrak{e}_{a,+b}),
(\mathfrak{e}^\dag_{a},\mathfrak{e}^\dag_{a,+b}) \right\} ,\;\;\;\;\;1\leq a< b\leq
4\;.
\end{eqnarray}
The first set generates $U(4)$, and the set of generators
$\{\mathfrak{e}_{ab}\}$ (resp.$\{\mathfrak{e}_{ab}^\dag\}$)  mutually commute. They satisfy the commutation relations
\begin{eqnarray}\label{commutation relations}
&&[\mathfrak{e}_{ab}\,,\,{\mathfrak{e}}_{cd}^\dag]
=\delta_{ac}\mathfrak{u}_{db}+\delta_{ad}\mathfrak{u}_{cb}
+\delta_{bc}\mathfrak{u}_{da}+\delta_{bd}\mathfrak{u}_{ca}\notag\\
&&[\mathfrak{u}_{ab}\,,\,{\mathfrak{u}}_{cd}]
=\delta_{bc}\mathfrak{u}_{ad}-\delta_{ad}\mathfrak{u}_{cb}\notag\\
&&[\mathfrak{u}_{ab}\,,\,{\mathfrak{e}}_{cd}]
=\delta_{bc}\mathfrak{e}_{ad}+\delta_{bd}\mathfrak{e}_{ac}\notag\\
&&[\mathfrak{u}_{ab}\,,\,{\mathfrak{e}}_{cd}^\dag]
=-\delta_{ac}{\mathfrak{e}}_{bd}^\dag
-\delta_{ad}{\mathfrak{e}}_{bc}^\dag\notag\\
&&[\mathfrak{e}_{ab}\,,\,{\mathfrak{e}}_{cd}]=
[{\mathfrak{e}}_{ab}^\dag\,,\,{\mathfrak{e}}_{cd}^\dag]=0\;.
\end{eqnarray}
Let $\varrho':\mathfrak{Sp}(8)\rightarrow L(\mathcal{V})$ be a complex representation
with $\mathcal{V}$ a $\mathfrak{G}$-module. The physically relevant
triangular decomposition of the algebra induces a decomposition of
$\mathcal{V}$ by
\begin{equation}\label{weight decomposition}
\mathcal{V}=\bigoplus_{w} \mathcal{V}_{(w)}
\,,\;\;\;\mathcal{V}_{(w)}:=\{\Bold{v}\in \mathcal{V}
:\varrho'(\mathfrak{h}_a)\Bold{v}=w_a\Bold{v}\}\,,\;\;\;a\in\{1,\ldots,4\}
\end{equation}
where $\mathfrak{h}_a\in\mathfrak{G}_0$ and ${w}=\{w_1,\ldots,w_4\}$
is a weight in the basis of fundamental weights
composed of eigenvalues $w_a\in\R$. It is well-known that a particular $\mathcal{V}$ can be generated by acting with raising operators $\mathfrak{g}_ +\in\mathfrak{G}_+$ on a dominant-integral lowest-weight vector $\Bold{v}_{w_-}$. Call this vector space $\mathcal{V}_{{w_-}}$.

Now, there is a distinguished subalgebra of $\mathfrak{G}$; its
maximal compact subalgebra $\mathfrak{U}(4)$. Let
$\mathcal{V}_{({\mu})}\subset\mathcal{V}_{{w_-}}$ denote the submodule
generated by $\mathfrak{U}(4)$ acting on the dominant-integral lowest-weight vector
$\Bold{v}_{w_-}$. The submodule $\mathcal{V}_{({\mu})}$ then furnishes an
irreducible representation (IR) of $U(4)$ where
${\mu}=[\mu_1,\ldots,\mu_4]$ is a partition based on $w_-$ that labels
the representation. Since $w_-$ is a lowest weight,
$\mathcal{V}_{({\mu})}$ is an invariant sub-space with respect to the
subalgebra $\mathfrak{P}:=\mathfrak{G}_-\cup\mathfrak{U}(4)$, i.e.
${\bar{\varrho}}'(\mathfrak{P})\mathcal{V}_{({\mu})}\subseteq\mathcal{V}_{({\mu})}$ where $\bar{\varrho}'$ is a restricted representation of $\varrho'$.
From this, one constructs representations of $\mathfrak{P}$ based on lowest-weight IRs of $U(4)$  and labeled by partitions $[\mu_1,\ldots,\mu_4]$.\footnote{Remind that $U(4)$ has \emph{both} boson and fermion representations.\cite[pg. 500]{G}}

Explicitly, the parabolic decomposition is described by
\begin{equation}
\mathfrak{Sp}(8)=\mathfrak{Z}_-\oplus\mathfrak{U}(4)
\oplus\mathfrak{Z}_+
\end{equation}
where $\mathfrak{Z}_+=\mathrm{span}\{\mathfrak{e}_{ab}\}$,
$\mathfrak{Z}_-=\mathrm{span}\{{\mathfrak{e}}_{ab}^\dag\}$ and
$\mathfrak{U}(4)=\mathrm{span}\{{\mathfrak{u}}_{ab}
\}$.
Choosing the lowest weight as opposed to the highest weight vector to generate $\mathcal{V}_{(\mu)}$ seems arbitrary from a physical standpoint. Therefore, at least in this paper, we will assume that the system enjoys the symmetry $\mathfrak{Z}_+\rightleftarrows\mathfrak{Z}_-$.

Our goal is quantization and, following the program proposed in \cite{LA1}, quantum kinematics is governed by the complexified group --- $Sp(8,\C)$ in this case. So complexify the algebra while maintaining the parabolic decomposition and consider the quotient algebra
\begin{equation}
\mathfrak{Z}_+:=\frac{\mathfrak{Sp}(8)}
{\mathfrak{P}}:=\frac{\mathfrak{Sp}(8)}
{\mathfrak{Z}_-\oplus\mathfrak{U}(4)}
\end{equation}
and its associated complex coset space $Z:=Sp\,(8,\C)/P^\C$ with $\mathrm{dim}_\C(Z)=10$.
Since the elements of $\mathfrak{Z}_+$ mutually commute, we will eventually interpret $Z$ as the system configuration space and the elements of $\mathfrak{P}$ as the generators of `external' and `internal' dynamics. Accordingly, we propose to associate dynamical variables parametrizing `external'
interactions with the coset space $Sp\,(8,\R)/U(4)$. This we interpret as a
$\mathrm{dim}_\R(Sp\,(8,\R)/U(4))=20$ manifold whose associated algebra generates a non-commutative phase space.

The complex coset space $Z$ furnishes both a convenient physical interpretation and the means to construct induced URs.

\subsubsection{Induced URs}
So far we have only considered the algebra $\mathfrak{Sp}(8)$. But what we need to find is \emph{unitary} representations of the (simply connected) group $Sp(8,\C)$. This is a non-compact group, and we can't simply exponentiate a representation of its algebra because relevant representations are generally infinite-dimensional in this case. The method we will use to construct URs relies on Mackey's theory of induced representations which has been thoroughly developed \cite{M1}--\cite{KP}. We give only an
outline of the steps for a quantum system invariant under the involution $\mathfrak{Z}_-\rightleftarrows\mathfrak{Z}_+$:
\begin{description}
\item[step 1:]Find the \emph{basic} dominant-integral lowest-weight
modules of $\mathfrak{Sp}(8)$.  There are four: $\{\mathcal{V}^{(0)}_1,\mathcal{V}^{(1)}_8,\mathcal{V}^{(2)}_{27},\mathcal{V}^{(3)}_{48},\mathcal{V}^{(4)}_{42}\}$ where the subscript denotes the dimension of the module and the superscript labels the representation. The trivial representation $\mathcal{V}^{(0)}_1$ will eventually be identified with the quantum vacuum. The defining module is $\mathcal{V}^{(1)}_8$, and the adjoint module is $\mathcal{V}^{(0)}_1\oplus\mathcal{V}^{(1)}_8\oplus\mathcal{V}^{(2)}_{27}$. Whether there are other relevant representations based on $\mathcal{V}^{(3)}_{48}$ and $\mathcal{V}^{(4)}_{42}$ is unclear, but there is no reason not to expect them.

\item[step 2:]For each relevant representation, identify the dominant-integral lowest-weight vector  and generate the $\mathfrak{P}$ invariant sub-space $\mathcal{V}_{(\mu)}\subset\mathcal{V}_{w_-}$ for all relevant
unitary IRs of $U(4)$ by acting on the dominant-integral lowest-weight vector $\Bold{v}_{w_-}$. $U(4)$ being compact,
its unitary IRs have finite dimension, and, since they are dominant-integral, the $\mathcal{V}_{(\mu)}$ posses a positive definite Hermitian inner product.

\item[step 3:]Recall that the action of $\mathfrak{P}$
leaves $\mathcal{V}_{(\mu)}$ invariant. So maximum efficiency
(associated with the impending induced representation) obtains
through the quotient $Sp\,(8,\C)/P^\C$ which utilizes the
finite-dimensional $\mathcal{V}_{(\mu)}$. Accordingly,  extend the relevant UIRs of $U(4)$ to $P^\C$.\footnote{The extension from $U(4)$ to $P$ is trivial because, as follows from the commutation relations, $\mathfrak{Z}_-$ annihilates the lowest weight element in $\mathcal{V}_{(\mu)}$. Subsequently, the representation on $P$ can always be extended to $P^\C$.} Since the ten elements
in the factor algebra $\mathfrak{Z}_+$ mutually
commute,  we can anticipate that they yield a basis for
compatible quantum observables.

\item[step 4:]Construct the principal coset bundle
$(\mathcal{P},Z_\p,\breve{pr},P^\C)$ and its associated vector bundle
$(\mathcal{V},Z_\p,pr,\mathcal{V}_{(\mu)},P^\C)$ where the base space
may be a submanifold with boundary of the homogeneous coset space $Z_\p\hookrightarrow Z:=Sp\,(8,\C)/P^\C$.
Recall that a point $g\in \mathcal{P}\equiv Sp\,(8,\C)$ is an admissible map
$g:\mathcal{V}_{(\mu)}\rightarrow\mathcal{V}$. Since we are
stipulating unitary IRs of $U(4)$, there is a unique (up to a scalar multiple) lowest weight
$\Bold{v}_{w_-}\in\mathcal{V}_{(\mu)}$ invariant under the right action of $P$ so that
$g(\Bold{v}_{w_-})$ can be identified with the zero-section in
$\mathcal{V}$. It is important that  the elements of
$\mathfrak{Z_+}$ mutually commute since then
$\mathrm{exp}\{\mathfrak{z}_+\}(\mathcal{V}_{(\mu)})$
induces a foliation of $\mathcal{V}$ compatible with the fiber
structure, i.e. leaves are homeomorphic to $Z$.

\item[step 5:]The usual construction of induced representations starts with the vector space of continuous, equivariant functions on $\mathcal{P}$. Here we assume without proof the construction can be extended to the space of continuous, equivariant, vector-valued $k$-forms on the $k$-tensor bundle $T^k(\mathcal{P})$.

For each relevant $\mathcal{V}^{(r)}_{(\mu)}$ (that is, $\mathcal{V}_{(\mu)}$ labeled by the  representation $r$ which is determined by $U(4)$ quantum numbers and the right action of $\mathfrak{Z}_-$ on $Z$), consider
continuous, compactly-supported, right-equivariant $\mathrm{k}$-forms $\breve{\psi}\in
C_C(T^k(\mathcal{P}),\mathcal{V}^{(r)}_{(\mu)})$ with norm
\begin{equation}\label{p-norm}
\|\breve{\psi}\|_{L^2} =\left(\mathrm{tr}\int_{Sp(8,\C)}
\breve{\psi}\wedge\star\,\breve{\psi}\right)^{1/2}\;<\infty
\end{equation}
where tr denotes the inner product on $\mathcal{V}^{(r)}_{(\mu)}$, equivariant means $R^\ast_{\tilde{g}}\breve{\psi}=\bar{\varrho}(\tilde{g}^{-1})\breve{\psi}$, and the Hodge $\star$ comes from the left-invariant volume form on $Sp(8,\C)$. The normalized induced representations are then furnished by
\begin{equation}\label{induction}
\mathbf{NE}_{P^\C}^{Sp\,(8,\C)^{(r)}}
=\{\breve{\psi}\in
L^2(T^k(\mathcal{P}),\mathcal{V}^{(r)}_{(\mu)})\,|\breve{\psi}(g\,p)
=N(p)\bar{\varrho}(p^{-1})\breve{\psi}(g)\}
\end{equation}
where $\breve{\psi}(gp)$ is shorthand for $\breve{\psi}(R'_p(g)\mathfrak{g}_1,\ldots,R'_p(g)\mathfrak{g}_k)$ with the derivative of right group action $R'_p(g):T_g(\mathcal{P})\rightarrow T_{gp}(\mathcal{P})$. The normalization constant is determined by $N^2(p):={\triangle_{P}(p)/
\triangle_{G}(p)}$ with
$\triangle_G(g)=|\mathrm{det}\,Ad_G(g)|$ the modular function of
$G$ and $\bar{\varrho}:P^\C\rightarrow
L_B(\mathcal{V}^{(r)}_{(\mu)})$ is a unitary lowest-weight IR.\footnote{Given a canonical section $s_i$, one can identify $\breve{\psi}$ with a section of $\Lambda^k\mathcal{P}\otimes\mathcal{V}^{(r)}_{(\mu)}$.\cite{LA2}}

\item[step 6:]Construct the Whitney sum bundle
\begin{eqnarray}
\mathcal{W}_{\mathcal{V}}&:=&(\bigoplus_r\mathcal{V}^{(r)},Z_\p,pr,
\bigoplus_r\mathcal{V}^{(r)}_{(\mu)},P^\C)\notag\\
&=:&\left(\mathcal{W},Z_\p,pr, \mathcal{W}_{(\Bold{\mu})},P^\C\right)
\end{eqnarray}
using all relevant unitary IR
modules $\mathcal{V}^{(r)}_{(\mu)}$. The typical fiber $\mathcal{W}_{(\Bold{\mu})}$ is Hilbert.

The representation furnishing module is
\begin{equation}
\mathbf{NE}_{P^\C}^{Sp\,(8,\C)}=\bigoplus_r\mathbf{NE}_{P^\C}^{Sp\,(8,\C)^{(r)}}\;.
\end{equation}
It can be given a Hilbert structure with inner product
\begin{equation}\label{first inner product}
\langle\breve{\psi}_1|\breve{\psi}_2\rangle_{\mathbf{NE}_{P}^{Sp\,(8,\C)}}
:=\mathrm{tr}\int_{Sp(8,\C)}
\breve{\psi}\wedge\star\,\breve{\psi}
=:\mathrm{tr}\int_{Sp(8,\C)}
(\breve{\psi}|\breve{\psi})\Bold{\tau}
\end{equation}
where the trace is now on $\mathcal{W}_{(\Bold{\mu})}$, the inner product on $\Lambda^k\mathcal{P}$ is $(\cdot|\cdot)$, and $\Bold{\tau}$ is the left-invariant volume element on $Sp(8,\C)$.  The continuous, unitary induced representation $\mathrm{ind}^{Sp(8,\C)}_{P^\C}\bar{\varrho}:Sp\,(8,\C)\rightarrow
L_B(\mathbf{NE}_{P^\C}^{Sp\,(8,\C)})$, which will \emph{not} be irreducible in general, can be expressed as
\begin{eqnarray}
(\mathrm{ind}^{Sp(8,\C)}_{P^\C}\bar{\varrho}(\tilde{g})\breve{\psi})({g})
=(L_{\tilde{g}^{-1}}^\ast\breve{\psi})({g})
&=&\breve{\psi}((L^{-1})'_{\tilde{g}}({g})\mathfrak{{g}}_1,
\ldots,(L^{-1})'_{\tilde{g}}({g})\mathfrak{{g}}_k)\notag\\
&=&\bar{\varrho}(\tilde{g})\breve{\psi}((L^{-1}_{\tilde{g}}R_{\tilde{g}})'({g})\mathfrak{{g}}_1,
\ldots,(L^{-1}_{\tilde{g}}R_{\tilde{g}})'(g)\mathfrak{{g}}_k)\notag\\
&=&\bar{\varrho}(\tilde{g})(Ad_{\tilde{g}^{-1}}^\ast\breve{\psi})(g)
\end{eqnarray}
where $\tilde{g},{g}\in Sp\,(8,\C)$ and $L$ is the left group action.
\end{description}

Unitarity of $\mathrm{ind}^{Sp(8,\C)}_{P^\C}\bar{\varrho}$ follows from the left invariance of $\Bold{\tau}$. To see it is an honest representation note that
\begin{eqnarray}
 (\mathrm{ind}^{Sp(8,\C)}_{P^\C}\bar{\varrho}(\tilde{g}_1\tilde{g}_2)\breve{\psi})({g})
 &=&(L^\ast_{(\tilde{g}_1\tilde{g}_2)^{-1}}\breve{\psi})({g})\notag\\
 &=&((L_{\tilde{g}_2^{-1}}\circ L_{\tilde{g}_1^{-1}})^\ast\breve{\psi})({g})\notag\\
 &=&(L_{\tilde{g}^{-1}_2}^\ast\breve{\psi})((L^{-1})'_{\tilde{g}_1}({g})\mathfrak{{g}}_1,
\ldots,(L^{-1})'_{\tilde{g}_1}({g})\mathfrak{{g}}_k)\notag\\
 &=&\left(\mathrm{ind}^{Sp(8,\C)}_{P^\C}\bar{\varrho}(\tilde{g}_1)\circ
 \mathrm{ind}^{Sp(8,\C)}_{P^\C}\bar{\varrho}(\tilde{g}_2)\breve{\psi}\right)({g})\;.
 \end{eqnarray}
 Furthermore, since $\bar{\varrho}$ is unitary, $\langle\breve{\psi}_1|\breve{\psi}_2\rangle_{\mathbf{NE}_{P}^{Sp\,(8,\C)}}$ is invariant under vertical automorphisms of $\mathcal{P}$. Insofar as a vertical automorphism can be viewed as a change of basis in $\mathcal{W}_{(\mu)}$ and the choice of a fiducial basis is arbitrary, we identify equivalence classes of $\breve{\psi}\in\mathbf{NE}_{P}^{Sp\,(8,\C)}$ under vertical automorphisms (i.e. gauge equivalent $\breve{\psi}$) with kinematical quantum states.

By postulate, $\mathcal{H}\equiv \mathbf{NE}_{P}^{Sp\,(8,\C)}$ models the kinematical quantum Hilbert space, and gauge equivalence classes of  state vectors represent physical states.
It must be emphasized that $\mathcal{H}$ does not necessarily coincide with the \emph{physical} Hilbert space $\mathcal{H}_D$ generated by the assumed dynamical group $G_D$.\cite{LA1} In our case, the dynamical group is $Sp(8,\R)$ so eventually a suitable sub-representation must be identified.

\begin{remark}\label{phase space induction} Alternatively, the induction game can be played on the intermediate
vector bundle $(\mathcal{I},Sp\,(8,\C)/U(4,\C),
pr,\mathcal{W}_{(\Bold{\mu})}\,,U(4,\C))$ where $Sp\,(8,\C)/U(4,\C)$ can be
interpreted as a complex phase space. Non-trivial cross sections exist
for vanishing Euler class, and a  $2$-form on the space of sections
can be defined in the usual way in terms of the non-degenerate
$2$-form on the base space. However, since the elements of
$\mathfrak{Z}:=\mathfrak{Z}_+\oplus\mathfrak{Z}_-$ do not all
commute, $\mathrm{exp}\{\mathfrak{z}\}(\mathcal{W}_{(\Bold{\mu})})$ does
not yield a compatible foliation of $\,\mathcal{I}$. Consequently, the
relationship between the principal bundle and associated vector bundle that allows
identification of the Hilbert space with the space of smooth cross
sections on the associated bundle no longer makes physical sense:\emph{\cite{LA1}}  Generically, more than
one cross section is associated with an orbit of
$\mathrm{exp}\{\mathfrak{z}\}$. To remedy the situation, it is enough
to pick any ten commuting elements in $\mathfrak{Z}$; thus defining
a ``polarization" on the phase space which in turn induces a compatible
foliation.
\end{remark}

\subsubsection{Coherent states}
For simplicity, \emph{from here on restrict $\breve{\psi}$ to vector-valued $0$-forms}.

The URs from the previous subsection are generally reducible so their physical interpretation is unclear. To make headway, it is helpful to reformulate the  Hilbert space in terms of sections of the associated vector bundle $\mathcal{W}_{\mathcal{V}}$.\cite[\S2.2]{LA1} One finds the Hilbert space $\mathcal{H}=L^2(Z,\mathcal{W})$ equipped with an inner product coming from $\mathcal{W}$ and a quasi-invariant, left Haar measure on $Z$ given by
\begin{equation}\label{inner product}
\langle\psi_1|\psi_2\rangle=\int_{Z_\p}(\psi_1(z)|\psi_2(z))_{\mathcal{W}_z}\;d\nu_{P^\C}(z)\;.
\end{equation}
where $\mathcal{W}_z$ is the Hilbert fiber over $z\in Z_\p$ and $\nu_{P^\C}(z)$ is a quasi-invariant measure on $Z$. Given a canonical section $s_i$ on $\mathcal{P}$, the corresponding canonical induced representation $\rho:Sp(8,\C)\rightarrow L_B(L^2(Z,\mathcal{W}))$ defined by $\rho({g})\circ s_i^\ast=s_a^\ast\circ\mathrm{ind}^{Sp(8,\C)}_{P^\C}\bar{\varrho}({g})$ is continuous, unitary, and invariant under vertical automorphisms of $\mathcal{P}$.\cite[\S2.2]{LA1}

The structure of the coset space
$Z:=Sp\,(8,\C)/P^\C$ immediately suggests defining Perelomov-type CS. Many useful details regarding these and other types of CS can be found in
\cite{DQ} and \cite{BR}.\footnote{Another important and useful type of CS is the Barut/Girardello-type. They are analogs of momentum eigenstates of Poincar\'{e}.}
Recall that $g\in G^\C$ can be viewed as an admissible map $g:\mathcal{W}_{(\Bold{\mu})}\rightarrow\mathcal{W}$. Given an open region $U_i\in Z_\p$, a local
trivialization $\{U_i,\varphi_i\}$ of the Whitney sum bundle,  and a local chart
$\phi:U_i\rightarrow\C^{10}$; a point $w\in
\pi^{-1}(U_i)\subset\mathcal{W}$ can be represented on
$\C^{10}\times\mathcal{W}_{(\Bold{\mu})}$ as
\begin{equation}
|\phi(z);\Bold{\mu}):=\left(\exp\left\{\sum_{a,b}{z}^\ast_{ab}\,\mathfrak{e}_{ab}\right\}\right)| \Bold{\mu})
\end{equation}
where the point $z\in U_i$ has coordinates
$\phi(z)={z}^\ast_{ab}\in \C^{10}$ with $a,b\in\{1,\ldots,4\}$ such that $a\leq b$, the vector $|\Bold{\mu})\in\mathcal{W}_{(\Bold{\mu})}$, and we used the coset decomposition to parametrize $g=\exp\{\frac{1}{2}
\sum_{a,b}{z}^\ast_{ab}\,\mathfrak{e}_{ab}\}\exp\{\mathfrak{p}^\C\}$.

To simplify notation a bit choose normal coordinates and write
$|\phi(z);\Bold{\mu})\rightarrow|z^\ast;\Bold{\mu})$. Then a state vector
$\psi\in\mathcal{H}_D$ can be modeled locally on
$U_i\times\mathcal{W}_{(\Bold{\mu})}$. Explicitly,
\begin{definition}
Given a local trivialization of the bundle
$\mathcal{W}_{\mathcal{M}}$, the CS model of a state vector
$\psi\in\mathcal{H}$ is defined by\footnote{This mixed bracket
notation is a bit strange: On one hand it should not be confused
with the Hilbert space inner product
$\langle\cdot\,|\cdot\,\rangle_{\mathcal{H}}$ or the inner product
$(\cdot\,|\cdot\,)_{\mathcal{W}_{(\Bold{\mu})}}$. On the other hand, it
emphasizes that the object it defines is a CS model of a
state vector. It must be kept in mind that the notation
$(\cdot\,|\cdot\,\rangle$ implicitly assumes a local trivialization,
and can be interpreted as the expression of a state vector in the
``$z\otimes \Bold{\mu}$ representation''.}
\begin{equation}
({z};\Bold{\mu}|\psi\rangle=:\Bold{\psi}_{\Bold{\mu}}(z) \equiv
s_i^\ast\breve{\psi}(z)
\end{equation}
where $z\in Z_\p\subseteq Z$ and $s_i$ is the canonical local section associated with the trivialization. The space $Z_\p$ is determined by boundary conditions on $\Bold{\psi}_{\Bold{\mu}}(z)$.
\end{definition}

We call $\Bold{\psi}_{\Bold{\mu}}(z)$ a coherent state wave function
or coherent state for short. It is a column vector according to
the relevant unitary IRs of $U(4)$ collectively labeled by
$\Bold{\mu}=(\mu^{(r_1)},\ldots,\mu^{(r_n)})$. Since
$\mathcal{V}^{(r)}_{w_-}$ are unitary IRs,
$\Bold{\psi}_{\Bold{\mu}}(z)$ is comprised of components
$\Bold{\psi}_{\Bold{\mu}}(z)
=(\Bold{\psi}^{r_1}_{\mu}(z),\ldots,
\Bold{\psi}^{r_n}_{\mu}(z))$ that do not mix --- a kind-of
super selection.  We will often restrict to a specific component and write
$\Bold{\psi}_{{\mu}}(z)=(z;\mu|\psi\rangle\in\mathcal{V}^{(r)}_{(\mu)}$ without indicating the representation $r$
for notational and conceptual simplicity. Because we have assumed $\mathfrak{Z}_-\rightleftarrows\mathfrak{Z}_+$ symmetry, $\Bold{\psi}_{\Bold{\mu}}(z)$ can be holomorphic or anti-holomorphic.

Similarly, operators have a coherent state representation:
\begin{definition}
The CS model of an operator ${O}\in L_B(\mathcal{H})$ is
defined by
\begin{equation}\label{CS operator}
({z};\Bold{\mu}|\,{O}\,\psi\rangle
=:\widehat{{O}}\,\Bold{\psi}_{\Bold{\mu}}(z)\;.
\end{equation}
\end{definition}

\subsubsection{Matrix CS}
The isomorphism between the space of $z_{ab}$ parameters and the
vector space of complex symmetric $4\times 4$ matrices allows to write the
coherent state basis as
\begin{equation}
|z^\ast;\Bold{\mu})=|\Bold{Z}^\ast;\Bold{\mu})
:=\left(\exp\left\{\frac{1}{2}
\mathrm{tr}(\Bold{{Z}}^\ast\mathfrak{E}_+)\right\}\right)| \Bold{\mu})
\end{equation}
where $\Bold{{Z}}^\ast\in M_4^{sym}(\C)$ is comprised of the coordinates
${z}^\ast_{ab}$ and it is understood that $U_i$ is modeled on
$M_4^{sym}(\C)$ --- the space of symmetric $4\times 4$ matrices with complex components.

To implement this, let $(\Bold{Z})_{ab}=(1+\delta_{ab})z_{ab}$ and define the symmetric matrices $\mathfrak{E}_+$
and $\mathfrak{E}_-$ with components  $\{\mathfrak{e}_{ab}\}$ and
$\{\mathfrak{e}^\dag_{ab}\}$ respectively, and $\mathfrak{E}_U$
comprised of $\{\mathfrak{u}_{ab}\}$. Form
the vector space $M_4^{sym}(\C)\otimes\mathcal{W}_{(\Bold{\mu})}$, and model $Z$ on
$M_4^{sym}(\C)$. Given a chart on $Z$ and a local trivialization $\{U_i,\varphi_i\}$ on
$\mathcal{W}$, a point $w\in\mathcal{W}$ is represented by
\begin{equation}
\varphi_i(w)=|\Bold{Z}^\ast;\Bold{\mu}) =\left(\exp\left\{\frac{1}{2}
\mathrm{tr}(\Bold{{Z}}^\ast\mathfrak{E}_+)\right\}\right)|\Bold{\mu})\;.
\end{equation}
Now define;
\begin{definition}
A CS model of a state vector in the matrix picture is defined by
\begin{equation}
({\Bold{Z}};\Bold{\mu}|\psi\rangle=(\Bold{\mu}|
\left(e^{\frac{1}{2}
\mathrm{tr}\,({\Bold{{Z}}}\mathfrak{E}_-)}\right)\,
\psi\rangle=:\Bold{\psi}_{\Bold{\mu}}(\Bold{Z})\;,
\end{equation}
and the model of an operator (not necessarily bounded) is
\begin{equation}
({\Bold{Z}};\Bold{\mu}|{O}\,\psi\rangle
=:\widehat{{O}}\,\Bold{\psi}_{\Bold{\mu}}(\Bold{Z})\;.
\end{equation}
\end{definition}

Referring to appendix A, an explicit CS realization of the Lie algebra
generators  for $0$-forms in a local trivialization
$U_i\times\mathcal{W}_{(\Bold{\mu})}$ in the matrix picture  is given by
\begin{equation}\label{generators1}
\widehat{\mathfrak{E}}_-=\partial_{\Bold{Z}}\otimes\Bold{I}_{(\Bold{\mu})}\;,
\end{equation}
\begin{equation}
 \widehat{\mathfrak{E}}_{+}
 =(\Bold{Z}\p_{\Bold{Z}}-5\Bold{I})\Bold{Z}\otimes I_{(\Bold{\mu})}-(\Bold{Z}\mathcal{U})_{sym}\;,
\end{equation}
and
\begin{equation}
 \widehat{\mathfrak{E}}_{U}
 =\Bold{Z}\p_{\Bold{Z}}\otimes I_{(\Bold{\mu})}+\mathcal{U}
\end{equation}
where $\Bold{Z}\mathcal{U}=(\Bold{Z}\Bold{U})_{\Bold{\mu}\Bold{\mu}'}$ and $(\Bold{A}\Bold{B})_{sym}\equiv1/2(\Bold{A}\Bold{B}+(\Bold{A}\Bold{B})^\mathrm{T})$ and $\mathcal{U}$ has matrix elements
\begin{equation}
\mathcal{U}_{ab,\Bold{\mu}'\,\Bold{\mu}}=(\Bold{U})_{\Bold{\mu}'\,\Bold{\mu}}
:=(\Bold{\mu}'|\bar{\varrho}'(\mathfrak{E}_{\,U})|
\Bold{\mu})_{\mathcal{W}_{(\Bold{\mu})}}
\end{equation}
with $\mathfrak{E}_{\,U}$  a $4\times 4$ matrix comprised of the generators of $U(4)$, i.e. $(\mathfrak{E}_{\,U})_{ab}=\mathfrak{u}_{ab}$. These Lie algebra generators do not generally belong to $L_B(\mathcal{H})$ but (restricted to a suitable domain) their unitary exponentials do.

In words, the
set of matrix-valued operators
$\{\widehat{\mathfrak{E}}_+,\widehat{\mathfrak{E}}_-,
\widehat{\mathfrak{E}}_U\}$ is a Perelomov-type CS model of the Lie algebra basis
$\{\mathfrak{e}_{ab},\mathfrak{e}^\dag_{ab},\mathfrak{u}_{ab}\}$ in the matrix
picture, and unitary exponentiation realizes an induced UR of $Sp\,(8,\C)$ near the identity. Of special note is the mixed $\Bold{Z},\mathcal{U}$ dependence of both $\widehat{\mathfrak{E}}_+$ and $\widehat{\mathfrak{E}}_U$.

\begin{remark}\label{Cayley remark}
Since $\mathcal{H}_D=Sp(8,\R)$ by assumption, a sub-representation must be identified. But defining CS directly in terms of the map $\exp:\mathfrak{Sp}(8)\rightarrow Sp(8,\R)$ is problematic because, although it makes sense in a local neighborhood of the identity element $e\in Sp(8,\R)$, it is not one-to-one or onto. A more sound approach is to define Cayley CS by $|\Bold{Z}^\ast;\Bold{\mu}):=\mathcal{C}( \Bold{Z}^\ast\mathfrak{E}_+)\,|\Bold{\mu})$  and conjugate CS by $|\Bold{Z};\Bold{\mu}):=\mathcal{C}^{\dag}( \Bold{Z}\mathfrak{E}_-)\,|\Bold{\mu})$ where $\mathcal{C}$ denotes the Cayley transform
 \begin{equation}
 \mathcal{C}( \Bold{Z}^\ast\mathfrak{E}_+)=(Id+\Bold{Z}^\ast\mathfrak{E}_+)(Id-\Bold{Z}^\ast\mathfrak{E}_+)^{-1}
\end{equation}
with $\Bold{Z}\Bold{Z}^\ast\neq\Bold{I}$. Then $|\Bold{Z}^\ast;\Bold{\mu})\oplus|\Bold{Z};\Bold{\mu})$ along with the anti-involution $J$ is a symplectic vector space, and $g\in Sp(8,\R)$ is represented by $\exp\{\pm\mathfrak{g}\}$ on each component sub-space(see for example \emph{\cite[pg. 18]{AN}}). So the original CS model remains intact, but there are two copies --- a conjugate pair. This is essentially a generalization of the Bargmann representation in elementary QM.

\end{remark}

The \emph{non-trivial} reproducing kernel for $(\Bold{Z}',\Bold{Z})\in U_i$
has been calculated explicitly by \cite{DQ} and reproduced in appendix A for convenience;
\begin{eqnarray}\label{overlap}
({\Bold{Z}'};\Bold{\mu}'|\Bold{Z}^\ast;\Bold{\mu})
=(\Bold{\mu}'|\rho\left(e^{\mathrm{tr}(\Bold{B}(\Bold{Z}',\Bold{Z}^\ast)\mathfrak{E}_U)}\right)|\Bold{\mu})
&=:&(\Bold{K}(\Bold{Z}',\Bold{Z}^\ast))_{\Bold{\mu}'\,\Bold{\mu}}
\end{eqnarray}
where $e^{-\Bold{B}(\Bold{Z}',\Bold{Z}^\ast)}:=(\Bold{I}-\Bold{Z}'\Bold{Z}^\ast)$ with $\Bold{Z}\Bold{Z}^\ast\in B_1^{open}(\Bold{I})$. (Recall that $\Bold{Z}\in M_4^{sym}(\C)$ implies $\Bold{Z}\Bold{Z}^\ast$ is positive-semidefinite Hermitian.) This is a key feature of $Sp(8,\C)$ due to $[\mathfrak{e}_{ab},\mathfrak{e}^\dag_{cd}]\neq0$ with important implications as we will see. The associated
resolution of the identity on $\mathcal{H}$ is
\begin{eqnarray}
{{Id}}
&=&\int_{U_i}|\Bold{Z}^\ast;\Bold{\mu})\; \mathbf{d\,\Bold{\sigma}}(z)\;(
{\Bold{Z}};\Bold{\mu}|
\end{eqnarray}
with measure
\begin{equation}
\mathbf{d\,\Bold{\sigma}}(z)
:=\mathcal{N}\Bold{K}^{-1}(\Bold{Z},\Bold{Z}^\ast)
\;d\nu_{P^\C}(z)=:\Bold{P}(\Bold{Z})\;d\nu_{P^\C}(z)
\end{equation}
where $\mathcal{N}$ is a normalization constant, $\Bold{K}(\Bold{Z}',\Bold{Z}^\ast)\Bold{K}^{-1}(\Bold{Z},\Bold{Z}^\ast)=\delta(\Bold{Z}',\Bold{Z})Id$, and
$\widehat{{Id}}\,\Bold{\psi}_{\Bold{\mu}}(\Bold{Z})=({Id}\,\psi)(\Bold{Z})$.

From these, one obtains the local CS superposition on $\mathcal{W}$;
\begin{eqnarray}
|\psi\rangle_i
&=&\int_{U_i}|\Bold{Z}^\ast;\Bold{\mu})\; \mathbf{d\,\Bold{\sigma}}(z)\;(
{\Bold{Z}};\Bold{\mu}|\psi\rangle
\end{eqnarray}
which must then be extended globally to $Z_\p$ with Dirichlet/Neumann boundary conditions on the boundary, e.g. $({\Bold{Z}};\Bold{\mu}|\psi\rangle|_{\p Z_\p}=\mathit{\Psi}_{\Bold{\mu}}$. Similarly,
\begin{eqnarray}\label{operator symbol}
\langle\psi|{O}\,\psi\rangle
&=&\int_{Z_\p}\Bold{\psi}^\dag_{\Bold{\mu}'}(\Bold{Z}')\Bold{P}(\Bold{Z}')\,
\widehat{{O}}\,\Bold{\psi}_{\Bold{\mu}'}(\Bold{Z}')\;d\nu_{P^\C}(z')\notag\\
&=&\int_{Z_\p}\int_{Z_\p}\Bold{\psi}^\dag_{\Bold{\mu}'}(\Bold{Z}')\Bold{P}(\Bold{Z}')\,
(\Bold{Z}';\Bold{\mu}'|O|\Bold{Z}^\ast;\Bold{\mu})\,\Bold{P}(\Bold{Z})\Bold{\psi}_{\Bold{\mu}}(\Bold{Z})\;
d\nu_{P^\C}(z,z')\,\notag\\
&=:&\int_{Z_\p}\int_{Z_\p}\Bold{\psi}^\dag_{\Bold{\mu}'}(\Bold{Z}')\,
\Bold{K}_{O}(\Bold{Z}',\Bold{Z}^\ast)\,\Bold{\psi}_{\Bold{\mu}}(\Bold{Z})\;
d\nu_{P^\C}(z,z')\;.
\end{eqnarray}
With suitable restrictions, $\widehat{{O}}\,\Bold{\psi}_{\Bold{\mu}'}(\Bold{Z}')=(\Bold{Z}';\Bold{\mu}|O\psi\rangle$ can be rendered a distribution, and  $\Bold{K}_{O}(\Bold{Z}',\Bold{Z}^\ast)$ (which can be given a functional integral representation) can be interpreted as the matrix CS model of the propagator associated with operator ${O}$.

\subsubsection{CS vacuum}

Let $\Bold{w}_-:=(w_-^{(r_1)},\ldots,w_-^{(r_n)})$
denote the collection of dominant-integral lowest weights. We define the ground state
by
$\breve{\psi}_0(g):=\Bold{v}_{\Bold{w}_-}\in\mathcal{W}_{(\Bold{\mu})}
\;\forall g\in Sp(8,\C)$. And we define a vacuum state to be the ground state of a UIR induced from the trivial partition $\Bold{\mu}=[\mu,\mu,\mu,\mu]$. In this case, $\mathcal{W}_{(\Bold{\mu})}\supset\mathcal{V}_1^{(0)}$ is irreducible and one-dimensional such that $(\rho({p})\breve{\psi}_0)(g)\propto\Bold{v}_{{\mu}}$ for all ${p}\in P^\C$ with $\Bold{v}_{{\mu}}\in\mathcal{V}_1^{(0)}$.
This suggests a natural definition of the CS model of a
vacuum state vector;
\begin{definition}
The CS model of a vacuum state vector $\varphi_0\in\mathcal{H}$ is
defined by
\begin{equation}
({z};\Bold{\mu}|\varphi_0\rangle=:\Bold{v}_{{\mu}}(z)
\equiv\Bold{v}_{\mu}\;\;\;\;\;\forall z\in Z\;
\end{equation}
such that $\langle\varphi_0|\varphi_0\rangle_{\mathcal{H}}=|\Bold{v}_{\mu}|$.
\end{definition}
The matrix CS model of the generators acting on a  ground state give $\widehat{\mathfrak{E}}_-\Bold{\psi}_{0}(\Bold{Z})=0$, $\widehat{\mathfrak{E}}_+\Bold{\psi}_{0}(\Bold{Z})=(\Bold{Z}\mathcal{U})_{sym}\Bold{\psi}_{0}(\Bold{Z})$, and $\widehat{\mathfrak{E}}_U\Bold{\psi}_{0}(\Bold{Z})=\mathcal{U}\Bold{\psi}_{0}(\Bold{Z})$. In particular, the vacuum state is annihilated by $\widehat{\mathfrak{E}}_-$ and invariant under $\widehat{\mathfrak{E}}_U$, but generally $\widehat{\mathfrak{E}}_+$ and $\widehat{\mathfrak{E}}_U$ can elevate ground states in any other representation $\mathcal{V}_{(\mu)}^{(r)}$ at any given point $\Bold{Z}$.

By definition, the vacuum furnishes the trivial one-dimensional representation of the parabolic subgroup $P^\C$ so
$\breve{\varphi}_0(gp^{-1})=\bar{\varrho}(p)\breve{\varphi}_0(g)
=\Bold{v}_{{\mu}}\,\forall g\in Sp(8,\C)$ and $p\in P^\C$. Hence, the VEV with respect to $\varphi_0$ of
any observable $\mathrm{O}\in\mathfrak{A}$ is automatically gauge invariant;
\begin{equation}
\langle\varphi_0|{O}\,\varphi_0\rangle=\langle \rho(p)\varphi_0|{O}\,\rho(p)\varphi_0\rangle
=\langle\varphi_0|\rho(p)^\dag{O}\rho(p)\,\varphi_0\rangle
=\langle\varphi_0|Ad(p){O}\,\varphi_0\rangle
\end{equation}
where $O\in L(\mathcal{H})$ is the (not necessarily bounded) linear operator representing the observable $\mathrm{O}\in\mathfrak{A}$.
Equivalently, if the trace exists,
\begin{equation}
\mathrm{tr}\,\Bold{K}_{O}(\Bold{Z}',\Bold{Z}^\ast)
=\mathrm{tr}\,\rho^{-1}(p)\Bold{K}_{O}(\Bold{Z}',\Bold{Z}^\ast)\rho(p)
=\mathrm{tr}\,Ad(p)\Bold{K}_{O}(\Bold{Z}',\Bold{Z}^\ast)\;.
\end{equation}
In fact, more generally
$\langle\psi_{gp}|{O}\,\psi_{gp}\rangle
=\langle\psi_{g}|{O}\,\psi_{g}\rangle$ since $\psi_{gp}$ and $\psi_{g}$ belong to the same quantum-state equivalence class. That is, expectation values of observables cannot distinguish elements of $\mathcal{W}_{(\Bold{\mu})}$ related through vertical automorphisms of $\mathcal{P}$. As discussed above, this is a reflection of the ambiguity of the basis in $\mathcal{W}_{(\Bold{\mu})}$. It is appropriate to call $P^\C$ a gauge group and $\Bold{v}_{{\mu}}$ a gauge-invariant vacuum.

\begin{remark}
Let's take stock of what we have so far. We start with a parabolic decomposition of the dynamical group $Sp(8,\R)$ and use this decomposition to construct induced representations of $Sp(8,\C)$. From the induced representations, the Hilbert space of state vectors $L^2( Z,\mathcal{W})$ or $\mathbf{NE}_{P^\C}^{Sp(8,\C)}$ is constructed. A complex coset space $Z$ admits construction of matrix CS, allowing to model $\psi$ by the `wave function' $\Bold{\psi}_{\Bold{\mu}}(\Bold{Z})$.\footnote{This of course is a very loose term since $\Bold{\psi}_{\Bold{\mu}}(\Bold{Z})$ is generally an element of $\mathcal{W}_{(\Bold{\mu})}$.} The wave function model yields explicit realizations of operators in a manner familiar from elementary QM: In particular we indicated expressions for the operators associated with the generators of $\mathfrak{Sp}(8)$ for vector-valued $0$-forms. Just like elementary QM, the wave function $\Bold{\psi}_{\Bold{\mu}}(\Bold{Z})$ can be interpreted as the state vector $\psi$ in the `CS representation' $|\Bold{Z}^\ast;\Bold{\mu})$.
\end{remark}

What remains is to construct a model of the algebra  $\mathfrak{A}$ and represent the dynamics. As discussed previously, both of these spring from the assumed dynamical group $Sp(8,\R)$. So eventually, we will have to restrict CS to a sub-representation of $Sp(8,\C)$.

\subsection{The $C^\ast$ algebra}
The material in this subsection is a quick summary of \cite{LA1} and \cite{LA2} which may be consulted for notation and more detail.

Since $Sp(8,\R)$ is supposed to be the `shadow' of the topological group $G$ of units of the quantum algebra $\mathfrak{A}$, it is reasonable to assume that (at least part of) $\mathfrak{A}$ can be modeled by functionals $\mathrm{F}:G^\C\rightarrow L_B(\mathcal{H})$. After all, we expect the (assumed) bracket structure of $\mathfrak{A}$ to be carried by $G_\lambda$. Therefore, the space of Mellin-integrable functionals comprised of $\mathrm{F}:G^\C\rightarrow L_B(\mathcal{H})$, which is a $C^\ast$-algebra when properly defined, should be a good model of $\mathfrak{A}$ --- in the sense that it contains any observable that one could hope to measure.

We use functional Mellin transforms to construct the $C^\ast$-algebra. Functional Mellin transforms are a particular type of functional integral defined in \cite{LA2}. Roughly, a functional integral is defined by a \emph{family} of integral operators $\mathrm{int}_\Lambda:\mathbf{F}(G_\Lambda)\rightarrow \mathfrak{B}$ where $\mathfrak{B}$ is a Banach algebra and $\mathbf{F}(G_\Lambda):=\mathbf{F}(G)|_{G_\Lambda}$ is a family of  spaces of integrable functions $f\in L^1(G_{\lambda},\mathfrak{B})$ for all $\lambda\in\Lambda$.\footnote{Besides specifying a locally compact topological group, individual $\lambda$ refer to specifications such as initial conditions, boundary conditions, constraints, etc. that characterize the quantum system under consideration.}  For the purposes of this paper, $G_\lambda=Sp(8,\R)$ and  $\mathfrak{B}\equiv L_B(\mathcal{H})$, but functional Mellin requires complexified $G^\C$.

So, to define functional Mellin transforms in the context of SQM, we start with the data $(G^\C, L_B(\mathcal{H}),G^\C_{\Lambda})$ where   $G^\C_\lambda=Sp(8,\C)$ and $\mathcal{H}$ is a UR module.
\begin{definition}\emph{(\cite{LA2})}\label{Mellin def.}
Let the map $\rho:G^\C_{\lambda}\rightarrow
L_B(\mathcal{H})$  be a continuous, injective homomorphism. Define continuous
functionals $\mathbf{F}(G^\C)\ni \mathrm{F}:G^\C\rightarrow L_B(\mathcal{H})$
equivariant under right-translations\footnote{This prescription is for left-invariant Haar measures. For right-invariant Haar measures impose equivariance under left-translations.} according to $\mathrm{F}(gh)=\mathrm{F}(g)\rho(h)$. Then the functional Mellin
transform
$\mathcal{M}_\lambda:\mathbf{F}(G^\C) \rightarrow L_B(\mathcal{H})$ is
defined by
\begin{equation}
\mathcal{M}_\lambda\left[\mathrm{F};\alpha\right]
:=\int_{G^\C}\mathrm{F}(gg^{\alpha})\;\mathcal{D}_\lambda g=\int_{G^\C}\mathrm{F}(g)\rho(g^{\alpha})\;\mathcal{D}_\lambda g
\end{equation}
where $\alpha\in\mathbb{S}\subset\C$, $g^\alpha:=\exp_G(\alpha\log_Gg)$ and $\mathrm{F}(g)\rho(g^{\alpha})\in
L_B(\mathcal{H})$ where the space of bounded linear operators $L_B(\mathcal{H})$ is given the strong topology. Denote the space of Mellin integrable functionals by $\mathbf{F}_{\mathbb{S}}(G^\C)$.
\end{definition}

Define a norm on $\mathbf{F}_{\mathbb{S}}(G^\C)$ by $\|\mathrm{F}\|_{\mathbb{S}}:=\mathrm{sup}_{\alpha,\lambda}\|\mathcal{M}_\lambda\left[\mathrm{F};\alpha\right]\|$.
Assume that $\mathbf{F}_{\mathbb{S}}(G^\C)$ can be completed with respect to $\|\mathrm{F}\|_{\mathbb{S}}$ or some other suitably defined norm. Then
\begin{proposition}\emph{(\cite{LA2} prop. 4.2)} For suitably restricted $\alpha$,
$\mathbf{F}_{\mathbb{S}}(G^\C)$ is a $C^\ast$-algebra when
 endowed with the involution $\mathrm{F}^\ast(g^{1+\alpha}):=\mathrm{F}(g^{-1-\alpha})^\ast\Delta(g^{-1})$
 and equipped with a suitable topology.
\end{proposition}

Our proposal is that $\mathbf{F}_{\mathbb{S}}(G^\C)$ models the $C^\ast$-algebra $\mathfrak{A}$ that characterizes the physical properties of a quantum system. Since $L_B(\mathcal{H})$ is non-commutative and $Sp(8,\C)$ non-abelian, only functional Mellin with $\alpha=1$ provides a $\ast$-representation of observables in $\mathbf{F}_{\mathbb{S}}(G^\C)$; that is, only $\mathcal{M}_\lambda[\,\cdot\,;1]=:\mathit{\Pi}^{(1)}_\lambda:\mathbf{F}_{\mathbb{S}}(G^\C)\rightarrow L_B(\mathcal{H})$ yields a $\ast$-representation.\cite[Prop. 4.6]{LA2} However, it is still useful to maintain the general Mellin transform setup because $\mathcal{M}_\lambda[\,\cdot\,;\alpha]$ will be a $\ast$-representation for all $\alpha\in\mathbb{S}$ when one wants to calculate functional traces and functional determinants.

We now have a means of representing an observable $\mathrm{F}\in\mathbf{F}_{\mathbb{S}}(G^\C)$ as an operator $\mathit{\Pi}_\lambda^{(1)}(\mathrm{F})\in L_B(\mathcal{H})$. The corresponding matrix CS model of said operator is $(\Bold{Z};\Bold{\mu}|\mathit{\Pi}_\lambda^{(1)}(\mathrm{F})\,{\psi}\rangle
=\widehat{\mathit{\Pi}_\lambda^{(1)}(\mathrm{F})}\Bold{\psi}_{\Bold{\mu}}(\Bold{Z})$. In particular, for $\mathrm{F}$ of the form $\mathrm{F}=\mathrm{E}^{-\mathrm{H}}$ with skew-adjoint $\mathrm{H}$, this becomes
\begin{equation}
(\Bold{Z};\Bold{\mu}|\mathit{\Pi}_\lambda^{(1)}(\mathrm{F})\,{\psi}\rangle
=e^{-f(\widehat{\mathfrak{G}_\lambda^\C})}\Bold{\psi}_{\Bold{\mu}}(\Bold{Z})
\end{equation}
where $f(\widehat{\mathfrak{G}_\lambda^\C})=\widehat{\mathit{\Pi}_\lambda^{(1)}(\mathrm{Log}\,\mathrm{F}^\ast)}
=\widehat{\mathit{\Pi}_\lambda^{(1)}(\mathrm{H})}=:H_\lambda$.\cite[\S3]{LA1} The associated matrix CS realization of the expectation is
\begin{eqnarray}
\langle\psi|\mathit{\Pi}_\lambda^{(1)}(\mathrm{F})\,\psi\rangle_{\lambda}
&=&\int_{Z_\p}\Bold{\psi}^\dag_{\Bold{\mu}'}(\Bold{Z}')\Bold{P}(\Bold{Z}')\,
\widehat{\mathit{\Pi}_\lambda^{(1)}(\mathrm{F})}\,\Bold{\psi}_{\Bold{\mu}'}(\Bold{Z}')\;d\mu_{P^\C}(z')\notag\\
&=:&\int_{Z_\p}\int_{Z_\p}\Bold{\psi}^\dag_{\Bold{\mu}'}(\Bold{Z}')\,
\Bold{K}_{F}(\Bold{Z}',\Bold{Z}^\ast)\,\Bold{\psi}_{\Bold{\mu}}(\Bold{Z})\;
d\mu_{P^\C}(z,z')\;.\notag\\
\end{eqnarray}

Gauge-invariant observables are characterized by $\mathrm{O}=\mathrm{Ad}(p)\mathrm{O}$ for all $p\in P$. The adjoint action on $\mathbf{F}_{\mathbb{S}}(G^\C)$ gets represented as an adjoint action $Ad(p)$ on $\mathit{\Pi}_\lambda^{(1)}(\mathbf{F}_{\mathbb{S}}(G^\C))$ so that gauge-invariant operators obey ${O^{-\alpha}_\lambda}={Ad}(p)O^{-\alpha}_\lambda=\rho(p)\,O^{-\alpha}_\lambda\,\rho(p^{-1})$. As an example of a gauge-invariant operator, suppose an observable $\mathrm{O}_P$ is left-equivariant (in addition to being right-equivariant) and a central function with respect to $P$, i.e. $\mathrm{O}_P(p\,gp^{-1})=\mathrm{O}_P(g)$ for all $p\in P$. It follows from the definition of functional Mellin that ${(O_P)}^{-\alpha}_\lambda
=\rho(p)\,{(O_P)}^{-\alpha}_\lambda\,\rho(p^{-1})$\;. At least one class of this type is not too hard to construct: let $\mathrm{O}_P(g)=\mathrm{E}^{-\log_{G^\C}g}=:\mathrm{E}^{-\gamma_i\mathfrak{g}_{\mathrm{U}}^i}$ with $\gamma_i\in\C$ and $\mathfrak{g}_{\mathrm{U}}^i\in\mathrm{U}(\mathfrak{G}^\C)$ where $\mathrm{U}(\mathfrak{G}^\C)$ is the universal enveloping algebra. If $[\gamma_i\mathfrak{g}_{\mathrm{U}}^i,\mathfrak{p}]=0$ for all $\mathfrak{p}\in\mathfrak{P}$, then $\mathrm{O}_P(g)$ is left- and right-equivariant as well as a central function.

\subsection{Automorphisms}
\subsubsection{Complex structures}\label{complex structures}
$Sp(8,\R)$ contains an \emph{inner} automorphism $j$ that is anti-involutive $j^2=-e$ (with $e$ the identity in $Sp(8,\R)$) and satisfies $g^\dag j g=j$ for all $g\in Sp(8,\R)$. At the Lie algebra level this relation becomes $\mathfrak{g}^\dag j=-j\,\mathfrak{g}$. Evidently, $j$ induces an adjoint action ${Ad}(j):\mathfrak{Sp}(8)_\pm\rightarrow \mathfrak{Sp}(8)_\mp$ by $\mathfrak{g}_\pm\mapsto j\,\mathfrak{g}_\pm j^{-1}=-\mathfrak{g}_\pm^\dag=-\mathfrak{g}_\mp$.

Having eigenvalues $\pm i$, the `complex structure' $j$ allows $\mathfrak{Sp}(8,\R)$ to be given the structure of the complex algebra $\mathfrak{Sp}(8,\C)$. This complex structure obviously extends to $\mathcal{V}_{(\mu)}$ via $\varrho'$ if it is not already complex. Consequently, for any complexified $\mathcal{V}^\C_{(\mu)}$, there exists a basis that diagonalizes $J:=\rho({j})$ and induces the decomposition $\mathcal{V}^\C_{(\mu)}=\mathcal{V}_{(\mu)}^{+}\oplus\mathcal{V}_{(\mu)}^{-}$ where
\begin{equation}
\mathcal{V}_{(\mu)}^{\pm}:=\left\{\Bold{v}\in\mathcal{V}^\C_{(\mu)}\,|\,J\Bold{v}=\pm i\Bold{v}\right\}
\;,\;\;\;\;\forall\Bold{v}\in\mathcal{V}^\C_{(\mu)}\;.
\end{equation}
Hence, $J$ provides a means to transfer objects formulated in the context of $Sp(8,\C)$ into objects relevant to $Sp(8,\R)$ and vice versa.

The automorphism $j$ serves another important purpose. Recall that $\mathfrak{Sp}(8,\R)$ is endowed with a non-degenerate, bi-linear, symmetric form $B$ --- the Cartan-Killing metric. Together with $j$, this defines a symplectic form by $\mathit{\Omega}(\cdot,\cdot):=B(\cdot,{Ad}(j)\cdot)$. So $\mathfrak{Sp}(8,\R)$ has the structure of a symplectic vector space. Moreover, the metric and symplectic structures on $\mathfrak{Sp}(8,\R)$ can be combined to construct a complex-valued bilinear form $h:\mathfrak{Sp}(8,\R)\times\mathfrak{Sp}(8,\R)\rightarrow\C$ by
\begin{equation}
h(\mathfrak{g}_1,\mathfrak{g}_2)=g(\mathfrak{g}_1,\mathfrak{g}_2)-i\mathit{\Omega}(\mathfrak{g}_1,\mathfrak{g}_2)
\,,\;\;\;\;\;\forall \mathfrak{g}_1,\mathfrak{g}_2\in\mathfrak{Sp}(8,\R)
\end{equation}
where $g$ is not necessarily the Cartan-Killing metric. It is evident that $h$ is a sesquilinear form on $\mathfrak{Sp}(8,\C)$ restricted to a real sub-space $\mathfrak{R}\subset\mathfrak{Sp}(8,\C)$ defined by $\mathfrak{R}\cap j\,\mathfrak{R}=\{0\}$.

In addition to the inner complex structure $j$, there are two \emph{outer} anti-involutions $k$ and $l$ that satisfy $g\,k=k\,g $ and $g\,l=l\,g$ for all $g\in Sp(8,\C)$ with $k^2=-e$ and $l^2=-e$. The four maps $e,k,j,l$ exhaust all (anti)involutive automorphisms of $Sp(8,\C)$. They are very special automorphisms because they endow $\mathfrak{Sp}(8,\C)$ with three independent complex structures. Moreover, $j\,k=-kj$, $j\,l=-lj$, $k\,l=-l\,k$, and $jkl=-e$. So the linear maps $\{Ad(e),Ad(j),Ad(k),Ad(l)\}$ generate a quaternion algebra that acts on $\mathfrak{Sp}(8,\C)$. However, none of them generates evolution: $e$ trivially commutes with everything, $j$ is not self-adjoint, and $k,l$ are outer automorphisms.

Nevertheless, we can interpret their actions on $\mathcal{V}^{(r)}_{(\mu)}$ and hence on $\mathcal{H}$. We have already seen that $j$ exchanges $\mathfrak{g}_\pm\rightleftarrows-\mathfrak{g}_\mp$. Since we identify $\pm$ elements with charged states in the adjoint representation, $Ad(j)$ is interpreted as `particle$\rightleftarrows$anti-particle'\footnote{The term `particle' is being used loosely here. It refers to the excitation in the Fock space $F_f$ associated with the creation/annihilation operators realizing $\mathfrak{Sp}(8)$.} exchange. To interpret $k$ and $l$, construct an explicit representation; for example the defining representation. Then $j,k,l$ can be represented by
\begin{equation}
J:=\rho(j)=\left(
             \begin{array}{cc}
               0 & \Bold{I}_{4\times4} \\
               -\Bold{I}_{4\times4} & 0 \\
             \end{array}
           \right)\;,
\end{equation}
\begin{equation}
K:=\rho(k)=i\left(
             \begin{array}{cc}
               0 & \Bold{I}_{4\times4} \\
               \Bold{I}_{4\times4} & 0 \\
             \end{array}
           \right)\;,
\end{equation}
and
\begin{equation}
L:=\rho(l)=i\left(
             \begin{array}{cc}
               \Bold{I}_{4\times4} & 0 \\
               0 & -\Bold{I}_{4\times4} \\
             \end{array}
           \right)\;.
\end{equation}

Evidently $Ad(k)$ can be interpreted as `particle$\rightleftarrows$anti-particle' without charge exchange. Finally, $\rho(l)$ affects the sign of the eigenvalues of stationary states. This suggests we interpret $Ad(l)$ as charge exchange with respect to evolution reversal. Finally,  we have already insisted that $\mathcal{H}$ is invariant under $\rho(j)$ because the highest and lowest weights were identified, but there is no reason to impose invariance under $\rho(k)$ and/or $\rho(l)$. On the other hand, invariance under $\rho(jkl)=-Id$ is assured.

\subsubsection{Evolution}
Following \cite{LA1}, quantum dynamics of a closed system is generated by a continuous, unitary inner automorphism $\mathrm{F}\mapsto \mathrm{Ad}(h(t))\mathrm{F}$ where $h(\R)\subset U(\mathbf{F}_{\mathbb{S}}(G^\C))$ is in the group of unitary units of $\mathbf{F}_{\mathbb{S}}(G^\C)$. Here we suppose that $h(t)$ is determined by
\begin{equation}
\frac{dh(t)}{dt}:=-\mathfrak{h}_{\mathrm{U}}(h(t))=-\mathfrak{h}_{\mathrm{U}}(t)h(t)\,,\;\;\;\;h(t_a)=e
\,,\;\;\;\;\;\;\;\;t\in\R
\end{equation}
where $\mathfrak{h}_{\mathrm{U}}(t)\in\mathrm{U}(\mathfrak{G}^\C)$ is skew-adjoint, $\mathrm{U}(\mathfrak{G}^\C)$ is the universal enveloping Lie algebra, and $e$ is the identity group element.

Let
\begin{equation}
\mathfrak{h}_{\mathrm{U}}(t)=\sum_i\alpha_i(t)\,\mathfrak{g}_{\mathrm{U}}^i
\end{equation}
where $\alpha_i(t)$ are real analytic functions and $\mathfrak{g}_{\mathrm{U}}^i\in\mathrm{U}(\mathfrak{G}^\C)$ are skew-adjoint. Then $h(t)$  can be written
\begin{equation}
 h(t)=e^{-\widetilde{\mathfrak{h}}(t)}:=e^{-\sum_i\beta_i(t)\,\mathfrak{g}_{\mathrm{U}}^i}
\end{equation}
where $\widetilde{\mathfrak{h}}(t)$ is determined by
\begin{equation}
\frac{d\widetilde{\mathfrak{h}}_{\mathrm{U}}(t)}{dt}
=\sum_{n=0}^\infty\frac{B_n}{n!}\,ad^n\left(\mathfrak{h}_{\mathrm{U}}(t)\right)
\widetilde{\mathfrak{h}}_{\mathrm{U}}(t)
\end{equation}
where $B_n$ are Bernoulli numbers and the map $ad^n\left(\mathfrak{h}_{\mathrm{U}}(t)\right)$ is defined recursively by $ad^0\left(\mathfrak{h}_{\mathrm{U}}(t)\right)
\widetilde{\mathfrak{h}}_{\mathrm{U}}(t):=\widetilde{\mathfrak{h}}_{\mathrm{U}}(t)$ and
$ad^n\left(\mathfrak{h}_{\mathrm{U}}(t)\right)
\widetilde{\mathfrak{h}}_{\mathrm{U}}(t):=ad^1\left(\mathfrak{h}_{\mathrm{U}}(t)\right)ad^{n-1}\left(\mathfrak{h}_{\mathrm{U}}(t)\right)
\widetilde{\mathfrak{h}}_{\mathrm{U}}(t)$.

Alternatively, take,
\begin{equation}\label{product form}
 h(t)=\prod_i\,e^{-\gamma_i(t)\,\mathfrak{g}_{\mathrm{U}}^i}
\end{equation}
where $\gamma_i(t)$ are related to $\alpha_i(t)$ through a system of nonlinear differential equations.\cite{WN} This form of $h(t)$ is particularly well-suited for parabolic decomposition and CS realizations: The adjoint action $\mathrm{F}\mapsto \mathrm{Ad}(h(t))\mathrm{F}$ on $\mathbf{F}_{\mathbb{S}}(G^\C)$ induces a continuous, time-dependent unitary inner automorphism on $L_B(\mathcal{H})$ through the $\ast$-representation
\begin{eqnarray}
\mathit{\Pi}^{(1)}_\lambda(\mathrm{F})\mapsto\mathit{\Pi}^{(1)}_\lambda(\mathrm{F}(t))
:=\mathit{\Pi}^{(1)}_{\lambda}(\mathrm{Ad}(h(t))\,\mathrm{F})
&=&\mathit{\Pi}^{(1)}_\lambda(h(t)^{-1})\left[\mathit{\Pi}^{(1)}_\lambda(\mathrm{F})\right]\mathit{\Pi}^{(1)}_\lambda(h(t))\notag\\
&=&Ad(h(t))\mathit{\Pi}^{(1)}_\lambda(\mathrm{F})
\end{eqnarray}
where $\mathit{\Pi}^{(1)}_\lambda(h(t))=\rho(h(t))=e^{-i\rho'(\widetilde{\mathfrak{h}}(t))}
=\prod_i\,e^{-i\gamma_i(t)\,\rho'(\mathfrak{g}_i)}$.\cite{LA1}

The evolution operator is defined by $U(t):=\mathit{\Pi}^{(1)}_\lambda(h(t))$ which suggests to define a Schr\"{o}dinger state vector ${\psi}(t):=U(t){\psi}$. Then transition amplitudes have Heisenberg \emph{and} Schr\"{o}dinger representations as usual
\begin{equation}
\langle{\phi}|\mathit{\Pi}^{(1)}_{\lambda}(\mathrm{F}(t)){\psi}\rangle
=\langle{\phi}|U(t)^{-1}\,\mathit{\Pi}^{(1)}_\lambda(\mathrm{F})\,U(t)|{\psi}\rangle
=\langle{\phi}(t)|\mathit{\Pi}^{(1)}_\lambda(\mathrm{F})|{\psi}(t)\rangle\;.
\end{equation}
Unitarity supplies the connection between the Heisenberg and Schr\"{o}dinger pictures
\begin{equation}
\langle{\phi}|\mathit{\Pi}^{(\alpha)}_{\lambda}(\mathrm{F}(t)){\psi}\rangle
=\langle{\phi}|U(t)^{-1}\,\mathit{\Pi}^{(\alpha)}_\lambda(\mathrm{F})\,U(t)|{\psi}\rangle
=\langle{\phi}(t)|\mathit{\Pi}^{(\alpha)}_\lambda(\mathrm{F})|{\psi}(t)\rangle
\;.\notag
\end{equation}

Define the $\ast$-homomorphism $\pi_z:L_B(L^2(Z,\mathcal{W}))\rightarrow L_B(\mathcal{W}_z)$ by
\begin{equation}\label{pi_z}
\pi_z(\mathit{\Pi}_\lambda^{(\alpha)}(\mathrm{F}))\psi(z):=
(\mathit{\Pi}_\lambda^{(\alpha)}(\mathrm{F})\psi)(z)\;\;\forall z\in Z\;.
\end{equation}
Then the time-dependent transition amplitudes in the Heisenberg and Schr\"{o}dinger pictures can be expressed as
\begin{eqnarray}
\langle{\phi}|\mathit{\Pi}^{(\alpha)}_{\lambda}(\mathrm{F}(t)){\psi}\rangle
&=&\int_{Z_\p}\big{(}{\phi}(z)|\pi_z(\mathit{\Pi}_\lambda^{(\alpha)}\mathrm{F}(t))
{\psi}(z)\big{)}_{\mathcal{W}_z}\;d\mu_{P_D}(z)\notag\\
&=&\int_{Z_\p}\big{(}{\phi}(z;t)|\pi_z(\mathit{\Pi}_\lambda^{(\alpha)}\mathrm{F})
{\psi}(z;t)\big{)}_{\mathcal{W}_{z}}\;d\mu_{P_D}(z)\;.\notag\\
\end{eqnarray}

The dynamics can be expressed in the CS model of a Schr\"{o}dinger state vector $\Bold{\psi}_{\Bold{\mu}}(z;t):=U(t)\Bold{\psi}_{\Bold{\mu}}(z)$ as expected
\begin{equation}
\frac{d{\Bold{\psi}_{\Bold{\mu}}}(z;t)}{dt}
=-\widehat{H(t)}\,\Bold{\psi}_{\Bold{\mu}}(z;t)\;\;\;\;\forall\, t\in\R\;\mathrm{and}\;z\in Z_\p
\end{equation}
where $H(t)=\rho'(\widetilde{\mathfrak{h}}(t))\in L(\mathcal{H})$ is skew-adjoint. The CS model of $H$ is a vector field on the group jet bundle restricted to $Z_\p$.

The supposition that dynamics is governed by \emph{inner} automorphisms has important consequences: Along with generating the dynamics of observables, $h(t)$ also induces an adjoint action on  $\rho(G^\C)$. The action represents evolution in $\mathrm{U}(\mathfrak{G}_{\lambda}^\C)$ because
\begin{equation}\label{group Heisenberg}
\frac{d\rho'(\mathfrak{g}_{\mathrm{U}}(t))}{dt}
=\left[\widetilde{H}(t),\rho'(\mathfrak{g}_{\mathrm{U}}(t))\right]\;.
\end{equation}
 So in particular, $h(t)$ effects a change $p\mapsto Ad(h(t))p=:{p}(t)$  for all $p\in P^\C$. But from (\ref{product form}) it is easy to see that there exist some $h(t)$ such that $p(t)\notin P^\C$. Hence, after evolution, $\psi(z{p}(t))$ is no longer necessarily an element of $\mathcal{W}_z$: If it happens that $\psi(z{p}(t))\notin \mathcal{W}_z$ at some time $t\in\R$, then it makes sense to define a new parabolic subgroup $\widetilde{P}^\C=Ad(h(t))P^\C$ (along with its associated ground states) if it is stable during an evolution epoch with $t\in[t_{a'},t_{b'}]\subset\R$. Evidently ground states depend on the evolution history of a quantum system. Moreover, $\widetilde{P}^\C$ induces a new coset space $\widetilde{Z}$ with its associated CS model. In this sense $Z$ evolves, and the physical interpretation of CS is time-dependent. In other words, the kinematics is time-dependent in general.

But what about the vacuum? By definition, the vacuum furnishes the trivial representation of $Sp(8,\C)$ with $\langle\varphi_0|\varphi_0\rangle_{\mathcal{H}}=|\Bold{v}_\mu|$. So, for the expectation of a unitary evolution (which is precipitated by an observation/measurement that induces the homomorphism $G^\C\rightarrow G^\C_{\lambda}$), the vacuum doesn't change.\footnote{However, we do not \emph{a priori} exclude the possibility of non-unitary evolution of  a  quantum system that has been perturbed by an \emph{external} agent. The perturbation is still dictated by some subgroup of units in $\mathbf{F}_{\mathbb{S}}(G^\C)$, but they need not be unitary. This may lead to a new vacuum $\Bold{v}_{\widetilde{\mu}}$ induced from a new degenerate partition $\Bold{\widetilde{\mu}}=[\widetilde{\mu},\widetilde{\mu},\widetilde{\mu},\widetilde{\mu}]$ coming from $\widetilde{P}^\C$. The vacuum module $\mathcal{W}_{(\Bold{\widetilde{\mu}})}$ remains one-dimensional, but now  $\langle\widetilde{\varphi}_0|\widetilde{\varphi}_0\rangle_{\mathcal{H}}
=|{\Bold{\widetilde{v}}_{\widetilde{\mu}}}|\neq|{{\Bold{v}}_\mu}|$.}

\begin{remark}This section was meant to outline the quantization of $Sp(8,\R)$, but much of the construction was actually built using $Sp(8,\C)$. This should not pose a problem as long as we are careful to restrict to real objects, subrepresentations, and/or subspaces at the appropriate places making use of the complex structure $J$. Of course there is a lot of work associated with these restrictions that we have skipped.

By assumption, $Sp(8,\R)$ is the dynamical group. But, being multiply connected, it cannot be modeled by ordinary representations on $\mathcal{H}$. For that we  need its double cover $\widetilde{Sp}(8,\R)=Mp(8,\R)$ along the lines discussed in remark \emph{\ref{Cayley remark}}. Then, extracting the relevant $Mp(8,\R)$ observables and subrepresentations from $\mathbf{F}_{\mathbb{S}}(Sp(8,\C))$ would lead to $Mp(8,\R)$ invariant transition elements. There are likely interesting consequences hiding in these details.
\end{remark}

\section{Some physical interpretations}
 Although there are still many aspects of the quantization that merit further investigation, we will move on to some physical interpretation of SQM.

\subsection{`Observed geometry'}\label{observed geometry}
Points in the coset space $Z$ are generated by the observables ${\mathfrak{e}}^\dag_{ab}$. By now it is clear that the spectra of the associated operators $\mathcal{Z}:=\sigma(\widehat{\mathfrak{e}}^\dag_{ab})$ can be thought of as coordinates of a configuration space characterizing the CS parameter space. Likewise, the spectra $\sigma(\widehat{\mathfrak{e}}^\dag_{ab})\times\sigma(\widehat{\mathfrak{e}}_{ab})$ can be viewed as coordinates of an associated complex cotangent bundle $T^\ast\mathcal{Z}$ with $\mathrm{dim}_\C(\mathcal{Z})=10$ (assuming the spectra form a topological space).

However, we have assumed the quantum system dynamics is generated not by $Sp(8,\C)$ but by $Sp(8,\R)$. So the relevant CS for dynamical systems must be a sub-representation parametrized by a sub-space $X\subset Z$ with $\mathrm{dim}_\R(X)=10$. This leads to an associated cotangent bundle $T^\ast\mathcal{X}$. We want to explore the geometry of this bundle. But first some motivation:

The geometry of $Sp(8,\R)$ clearly transfers  to $T^\ast\mathcal{X}$ through its Lie algebra --- but in a guise that depends on the nature of the CS Hilbert space. Given a CS model of the group Lie algebra operators, we can expect to find symmetric and anti-symmetric forms associated with metric and symplectic forms on $\mathfrak{Sp}(8)$ as well as complex structures associated with the automorphisms $i,j,k$ (discussed in $\S$ \ref{complex structures}) that depend on the particular system under consideration. The primary task is to determine if this geometric space contains the seeds of a 4-d space-time. Observe that  $[\mathfrak{e}_{a},\mathfrak{e}_{a}^\dag]\sim\mathfrak{h}_a$ and $\mathfrak{h}_a$ generates the maximal abelian, rank-4 subgroup. For the real symplectic group, eigenvalues of these subgroup elements are purely unitary or purely real.  Accordingly, the topology of the
maximal abelian subgroup embedded in the group manifold $\mathbb{M}[Sp(8,\R)]$ is locally
$\mathbb{T}^{k}\times\R^{k'}$ where $k+k'=4$.\footnote{The group manifold of $Sp\,(8,\R)$ comprises five domains locally characterized by the five different topologies of the abelian subgroup manifold. Essentially this comes from the indefinite Killing forms on $\mathfrak{Sp}(8,\R)$ inherited from the Hermitian form on $\mathfrak{Sp}(8,\C)$. A thorough discussion of this and other aspects of evolution on non-compact group manifolds can be found in \cite{KM}.} It is evident that the ground-state inner product (GIP) in $\mathcal{W}_z$ of composite operators that generate the maximal abelian subgroup will be parametrized by $T(\mathbb{T}^{k}\times\R^{k'})\simeq\R^{k,k'}$.  So if we use a GIP to construct a non-degenerate symmetric quadratic form using the commutator subalgebra $[\mathfrak{e}_a,\mathfrak{e}_{-a}]$, we will have a model of a metric with signature $(k,k')$.
Similar reasoning applied to $[Ad(j)\mathfrak{e}_{a},Ad(j)\mathfrak{e}_{b}^\dag]$ and $[\mathfrak{e}_{a,b},\mathfrak{e}_{c,d}^\dag]$ implies we can construct a model of an almost complex structure and pre-symplectic form to go along with the metric. Ultimately, these stem from an $Ad$-invariant Hermitian form on $\mathfrak{Sp}(8,\C)$.

These properties, together with the observation that  $e$ and $j$ are the only two \emph{inner} involutive automorphisms,  motivate the definition of pre-geometry.
\begin{definition}\label{STA}
Let $R_{k,k'}$ represent the region in
$Sp\,(8,\R)$ whose maximal abelian subgroup is locally $\mathbb{T}^{k}\times\R^{k'}$ with $k\in\{0,\ldots,4\}$, and choose an
open region $U_i\subseteq R_{k,k'}$. The pre-geometry
$\mathcal{G}(k,k')\subset L(\mathcal{H})$ is generated by the image under $\rho'_\R$ of $\mathfrak{e}^\dag_{ab}$ (identified with the associated ten left-invariant vector fields
on $U_i$) together with their inner involutive automorphisms, i.e. for $E:=\rho_\R(e)=Id$, $E^\dag_{a}:=\rho'_\R(\mathfrak{e}^\dag_a)$,
$E^\dag_{a,b}:=\rho'_\R(\mathfrak{e}^\dag_{a,b})$, and
$J:=\rho_\R(j)$,
\begin{equation}
\mathcal{G}(k,k'):=\mathrm{span}_\R\left\{E,E^\dag_{a}, E^\dag_{a,b},
E^\dag_{-a},J\right\},\;\;\;\;k+k'=4
\end{equation}
where $a\neq b\in\{1,2,3,4\}$.
\end{definition}

Define $\mathit{\Pi}_a:=\rho'_\R(\mathfrak{e}^\dag_a-\mathfrak{e}_a)$ and $\mathit{\Pi}_{a,b}:=\rho'_\R(\mathfrak{e}^\dag_{a,b}-\mathfrak{e}_{a,b})$, and denote the collection by $\mathit{\Pi}_{i}\equiv\{\mathit{\Pi}_a,\mathit{\Pi}_{b,c}\}$ with $i\in\{1,\ldots,10\}$. Consider the $z$-dependent GIP
\begin{equation}
\frac{\big{(}\psi^{(r)}_0(z)\big{|}\pi'_z\{\mathit{\Pi}_{i},\mathit{\Pi}_{j}^\dag\}\,\psi^{(r)}_0(z)\big{)}_{\mathcal{W}_z}}
{(\psi^{(r)}_0(z)|\psi^{(r)}_0(z))_{\mathcal{W}_z}}
=:(g_{ij}^{(r)})_z
\end{equation}
where $\pi_z$ is defined in (\ref{pi_z}) and $\psi_0^{(r)}$ is a ground state and $\{\cdot,\cdot\}$ represents the Jordan product on $L(\mathcal{H})$. The definition is valid for ground states in each $\mathcal{V}^{(r)}_{(\mu)}$, and since $\mathcal{W}_{(\Bold{\mu})}$ is a direct sum of $\mathcal{V}^{(r)}_{(\mu)}$ the definition holds for the total ground state $\Bold{v}_{\Bold{w}_-}\in\mathcal{W}_{(\Bold{\mu})}$ as well.   In particular, if the total ground state happens to be the vacuum $\varphi_0$, then the vacuum-state inner product (VIP) $({g_{ij}^{(0)}})_z$ is $Sp\,(8,\R)$ invariant. Observe that the pre-geometry $\mathcal{G}(k,k')$ is an algebra  when restricted to the VIP, because $\varphi_0$ is a $U(4)$ singlet.\footnote{In this case, the algebra is closely
related but strictly different from what is called geometric algebra
in the literature. Specifically, $E_{a,b}$ is not the antisymmetric product of the $E_{a}$.} However, contrary to the vacuum case, $\mathcal{G}(k,k')$ is no longer an algebra with respect to a generic GIP.

For a non-trivial evolution of the total ground state,
\begin{eqnarray}
(g_{ij})_z(t)
&:=&\frac{\big{(}\psi_0(z)\big{|}U^{-1}(t)\pi'_z\{\mathit{\Pi}_{i},\mathit{\Pi}_{j}^\dag\}\,U(t)
\,\psi_0(z)\big{)}_{\mathcal{W}_z}}
{(\psi_0(z)|\psi_0(z))_{\mathcal{W}_z}}\notag\\
&=&\frac{\big{(}\psi(z;t)\big{|}\pi'_z\{\mathit{\Pi}_{i},\mathit{\Pi}_{j}^\dag\}
\,\psi(z;t)\big{)}_{\mathcal{W}_z}}
{(\psi_0(z)|\psi_0(z))_{\mathcal{W}_z}}\;;
\end{eqnarray}
and so this matrix is dynamical unless the ground state coincides with the vacuum. Similarly, non-trivial dynamics induces an anti-symmetric matrix
\begin{equation}
(\mathit{\Omega}_{ij})_z(t)
:=\frac{\big{(}\psi(z;t)\big{|}\pi'_z[\mathit{\Pi}_{i},\mathit{\Pi}_{j}^\dag]
\,\psi(z;t)\big{)}_{\mathcal{W}_z}}
{(\psi_0(z)|\psi_0(z))_{\mathcal{W}_z}}
\end{equation}
and an almost complex structure
\begin{equation}
({J})_z(t)
:=\frac{\big{(}\psi(z;t)\big{|}\pi_z(J)\,\psi(z;t)\big{)}_{\mathcal{W}_z}}
{(\psi_0(z)|\psi_0(z))_{\mathcal{W}_z}}\;.
\end{equation}

Consider the principal bundle $(\sigma(\mathcal{P}),\mathcal{Z},\breve{pr},P^\C)$ and its associated vector bundle where $\sigma(\mathcal{P})$ denotes the spectrum (with respect to a CS ground state) of the set of observables tangent to points in $Sp(8,\C)$.  Being associated with pre-geometry by definition, the GIPs of $\mathit{\Pi}_{a}$ and $\mathit{\Pi}_{a,b}$ are vector fields on $\mathcal{Z}$. Hence, the symmetric matrix $g_{ij}(t)$ corresponds to an evolution-dependent metric $g(\mathbf{v},\mathbf{w})(t):=\mathrm{v}^ig_{ij}(t)\mathrm{w}^j$ on $T\mathcal{Z}$ (assuming it is non-degenerate). Likewise, a non-degenerate $\mathit{\Omega}_{ij}(t)$ corresponds to a symplectic form. Together these define a Hermitian inner product $h_{ij}(t)$ on $T\mathcal{Z}$, and the almost complex structure allows identification of real sub-spaces. As posited, the relevant representation for dynamics is a real CS sub-representation so we expect $T^\ast\mathcal{X}\subset T^\ast\mathcal{Z}$ with $\mathrm{dim}_\R(\mathcal{X})=10$  where ${J}(t)$ provides the means to relate the two.

Evidently, for dynamical $Sp(8,\R)$, the GIPs of  pre-geometry characterize the geometry of an evolution-dependent cotangent bundle (assuming it remains a topological space under evolution), that can be interpreted as a phase-space. For want of a better name, we will call it the `observed phase space'.\footnote{In the limit of large systems such that $N=\mathrm{dim}_\C(\mathcal{W}_{(\Bold{\mu})})\rightarrow\infty$, `observed phase space' is posited to become classical, but a general phase-space would certainly not look classical for all systems. Note that $g$ and $\mathit{\Omega}$ are not related through ${J}$ so the observed
geometry is not K\"{a}hler in general.} To see this, note that these expectations, which encode the geometry, can only depend parametrically on the spectra $\mathcal{Z}$ because $\rho'(\mathfrak{p})\psi_0$ only transforms the $|\Bold{\mu})$ component of $\Bold{\psi}_0(z)$. Indeed, $\rho'(\mathfrak{e}^\dag_{a,b})$ annihilates the ground state, $\rho'(\{\mathfrak{u}_{ab}\})$ is unitary, and $\rho'([\mathfrak{e}_a,\mathfrak{e}^\dag_b])$ is normal. Hence, $Ad(p)\rho'([\mathfrak{e}_a,\mathfrak{e}^\dag_b])$ can be diagonalized by a unitary similarity transformation on $\mathcal{W}_{(\Bold{\mu})}$, which (once again) implies that a gauge transformation just corresponds to a change of coordinate basis in $\mathcal{W}_{(\Bold{\mu})}$.

Accordingly, observed phase space has $\mathrm{dim}_\R(T^\ast\mathcal{X})=20$ and the associated CS model of the operators $\widehat{\mathfrak{e}}_{ab}^\dag(t):=Ad(U(t))\mathfrak{e}_{ab}^\dag$ encode the time-dependence of the geometry through their CS ground-state spectra. Moreover, since
$\psi_{gp}\sim\psi_g$ for all $p\in P^\C$, the
geometry  is clearly gauge invariant for GIPs.

Now, for dynamical systems that stay near the ground state, one might expect the influence of the six elements $\mathfrak{e}^\dag_{a,b}$ on the phase-space geometry to be small if sizable GIPs of $E_{a,b}$ require large quantum numbers or, perhaps, large $\mathrm{dim}(\mathcal{W}_{(\Bold{\mu})})$ coming from tensor products of representations. In
this case, dynamics presumably would generate states with non-trivial support only on the diagonal of $\Bold{Z}$, and it would make sense to integrate the pre-geometric
structures over off-diagonal variables or simply restrict to diagonal variables to obtain an effective description. Recall $[\mathfrak{e}_a,\mathfrak{e}_a^\dag]\sim\mathfrak{h}$, so we can anticipate such dynamics to render real \emph{diagonal} sub-spaces of $\mathcal{Z}$ with metric signatures $(k,k')$.

To make this explicit, assume $\mathcal{Z}=\sigma(\widehat{\mathfrak{e}}_{ab}^\dag)$ is a topological space and let $\mathbb{M}^\C\subset \mathcal{Z}$ denote the $\mathrm{dim}_\C(\mathbb{M}^\C)=4$  sub-space associated with the diagonal elements in $\Bold{Z}$. Choose a pre-geometry $\mathcal{G}(k,k')$, and let $h_{ab}(m):=h_{ij}|_{T\mathbb{M}^\C}$ denote the associated Hermitian inner product on $T\mathcal{Z}$ restricted to $T\mathbb{M}^\C$. The restricted almost complex structure  $J(m):=J|_{T\mathbb{M}^\C}$  determines a real  sub-space $\mathbb{M}\subset \mathbb{M}^\C$. Evidently, $\mathbb{M}$ and $g_{ab}(m):=\Re[h_{ij}|_{T\mathbb{M}}]$
model a real manifold equipped with a metric of signature $(k,k')$ coming from the pre-geometry.\footnote{Essentially, $Z$ as a complex coset space possesses real subspaces with indefinite metric. One could alternatively use $\mathit{\Omega}_{ij}|_{T\mathbb{M}^\C}$ and $J(m)$ to define a real lagrangian subspace $\mathbb{L}\subset \mathbb{M}^\C$ and construct a metric $\tilde{g}_{ab}(l):=(J\mathit{\Omega}_{ij})|_{T\mathbb{L}}$.}

In this effective description,
$(\varphi_{\sim0}|\pi'_z\{\mathit{\Pi}_{a},\mathit{\Pi}^\dag_{b}\}|\varphi_{\sim0})_{\mathrm{diag}}$
along with pre-geometry $\mathcal{G}(1,3)$ (or $\mathcal{G}(3,1)$) could be interpreted as a model of an `observed
$4$-d space-time' with approximate Poincar\'{e}
symmetry  if the ground state remains very ``near''  the vacuum. (By ``near'' we mean the evolving ground state remains approximately a $U(4)$ singlet invariant under $\widehat{\mathfrak{E}}_-$.) Granted this perspective, Minkowski space-time and its concomitant irreducible representations owe their existence  to dynamically-induced approximate VIPs and so, in this sense, are emergent phenomena.

On the other hand, non-negligible GIPs of the $E_{a,b}$ can be interpreted as directed-area elements\footnote{One can utilize the defining representation of $\mathfrak{Sp}(8)$ described in appendix A and a computer to gain some intuition regarding the commutation relations among the $E_{ab}$.} according to the structure of the pre-geometry: So it
may be that physics based on $4$-d space-time is actually a
truncation of a $10$-d model
--- one in which both $g_{ij}$ and $\mathit{\Omega}_{ij}$ participate independently. It goes without saying that compactifying the extra six dimensions is neither necessary nor appropriate. For meso/macroscopic systems,
it is not hard to imagine that $E_{a,b}$ operators can have
significant expectation, and one can see the germ of vortex-type
dynamics that are algebraically independent from linear-type
dynamics.

\subsection{Particles and fields}

The fact that $P^\C$ generates $\mathcal{V}^{(r)}_{({\mu})}$ and the requirement that physical states are covariant under right-translation by $P^\C$ motivate the interpretation of `particles' and `fields':
\begin{definition}
An elementary particle is defined to be an eigenvector $\Bold{v}_{\Bold{p}}\in\mathcal{V}^{(r)}_{({\mu})}$ of $\rho'(\mathfrak{h}_a)$ with eigenvalues $\lambda_a$ for $a\in\{1,2,3,4\}$. In the CS model, an elementary field at a point $z\in Z_\p$ is defined by
\begin{equation}
\Bold{\psi}_{\Bold{v}_{\Bold{p}}}(z):=(z;\Bold{v}_{\Bold{p}}|\psi\rangle
=\int_{Z_\p}(z;\Bold{v}_{\Bold{p}}|z'^\ast;\Bold{\mu})\mathbf{d}\Bold{\sigma}(z')
\Bold{\psi}_{\Bold{\mu}}(z')
\end{equation}
and satisfies $\widehat{\rho'(\mathfrak{h}_a)}\Bold{\psi}_{\Bold{v}_{\Bold{p}}}(z)
=\lambda_{a}\,\Bold{\psi}_{\Bold{v}_{\Bold{p}}}(z)$. We call $\lambda_a$ (suitably normalized) $U(4)$ `charges'.
\end{definition}
Assuming suitable conditions, this can be inverted
\begin{equation}
\Bold{\psi}_{\Bold{\mu}}(z)
=\int_{Z_\p}(z;\Bold{\mu}|z'^\ast;\Bold{v}_{\Bold{p}})\mathbf{d}\Bold{\sigma}(z')
\Bold{\psi}_{\Bold{v}_{\Bold{p}}}(z')\;,
\end{equation}
and the wave function $\Bold{\psi}_{\Bold{\mu}}(z)$ can be interpreted as a superposition of elementary fields. Hence, the CS model of a state vector is a field by definition; albeit not necessarily elementary and definitely not a local operator.\footnote{We should emphasize this definition of field --- even at the operator level --- does not coincide with the notion of field in QFT. In QFT, a field is a superposition of creation and/or annihilation operators $c_\alpha, c^\dag_\alpha$, but in SQM there is no object to directly compare since everything is constructed from creation/annihilation \emph{products} contained in $c_{\alpha,\beta}$. To compare indirectly, the commutator of QFT fields roughly corresponds to the CS model of the corresponding SQM propagator.}

This is an obvious definition based on (\ref{weight decomposition}). That is, elementary particles span the full module $\mathcal{V}$ of relevant representations as required.  But --- contrary to the weight decomposition of $\mathcal{V}$ which leads to the (infinite-dimensional) irreducible discrete-series representations and, hence, particles characterized solely by their $U(4)$ partition --- the dominant-integral lowest-weight parabolic decomposition characterizes particles by their $U(4)$ charges \emph{and} their $P^\C$-induced representation. And, according to remark \ref{Cayley remark}, the conjugate CS model represents anti-particles.

 Notice the definition holds for all relevant representations labeled by $r$. Since $\rho'(\mathfrak{h}^2_a)$ can be interpreted as a number operator, multi-particle states are accounted for by appropriate $\mathcal{V}^{(r)}_{({\mu})}$ associated with tensor products and direct sums. For example, by definition the vacuum is an elementary CS field coming from the degenerate $U(4)$ representation labeled by some partition $\Bold{\mu}=[\mu,\mu,\mu,\mu]$. Since there is no \emph{a priori} reason to settle on a special value for $\mu$, it is not hard to imagine that different closed dynamical systems could have different vacua. After the vacuum, the next simplest IR is the defining representation of $U(4)$. It is standard to identify this IR with matter particles/fields.
Moving on to the adjoint representation, the CS model of state vectors in the adjoint representation can be interpreted as gauge fields. Gauge bosons\footnote{Note that, by our definition, gauge bosons are not elementary: they `live' in a tensor sum module of basic representations $\mathcal{V}_1^{(0)}\oplus\mathcal{V}_8^{(1)}\oplus\mathcal{V}_{27}^{(2)}$, and there is no reason (at this point at least) to assume the elementary particles associated with these representation spaces remain in proper relation to constitute eigenfields of $\mathfrak{h}_a$ at all $z\in Z$ under all circumstances.} are then naturally identified with eigenvectors $\mathfrak{p}\in\mathfrak{P}$ of $\mathfrak{h}_a$  in the adjoint representation. As the rank of $Sp\,(8,\R)$ is four, there are $26$ elementary gauge bosons that are characterized by four types of charge. Consequently, in the adjoint representation $\{\pi'_z(\rho'(\mathfrak{P}))\}$ correspond to gauge potentials on $Z$.\footnote{These are the closest analog to fields in QFT. Presumably, one could use these operators along with $\{\pi'_z(\rho'(\mathfrak{Z}_+))\}$ to make contact with gauge field theory.} It is remarkable that these gauge potentials posses both external and internal structure, and in a CS realization each component is a
matrix-valued differential operator whose dimension is dictated by the $\mathcal{W}_{(\Bold{\mu})}$. Finally, the last two basic representations $\mathcal{V}_{48}^{(3)}$ and $\mathcal{V}_{42}^{(4)}$ are suspected to be relevant, but their physical meaning as `elementary' is unclear. Observe, however, that the adjoint representation is disjoint from these two vector spaces.  In light of the absence of gauge boson interactions and since SQM dynamics can't mix representations, one wonders if $\mathcal{V}_{48}^{(3)}\oplus\mathcal{V}_{42}^{(4)}$ has dark matter implications.

\subsection{Matrix quantum mechanics}
Conspicuously absent from the physical interpretation thus far are the generators of $\mathfrak{Z}_+$. This of course is due to the fact that their role is to parametrize CS via $\Bold{Z}$. Nevertheless, \emph{for time-independent} $\widetilde{H}$, (\ref{group Heisenberg}) implies the $\mathfrak{Z}_+$ generators evolve according to the second-order operator equation
\begin{equation}\label{second-order}
\frac{d^2E_{ab}(t)}{dt^2}+ad^{\,2}(\widetilde{H})E_{ab}(t)=0
\end{equation}
where $E_{ab}:=\rho'(\mathfrak{e}_{ab})\in L(\mathcal{H})$ are a set of ten \emph{commuting} operators.

Since $E_{ab}(t)$ mutually commute, consider the eigenstates $E_{ab}(t)\mathit{\Psi}^{(\Bold{\lambda})}=\lambda_{ab}(t)\mathit{\Psi}^{(\Bold{\lambda})}$. In the CS model, this becomes
\begin{equation}
\widehat{E}_{ab}(t)\Bold{\Psi}_{\Bold{\mu}}^{(\Bold{\lambda})}(z)
=\lambda_{ab}(t)\Bold{\Psi}_{\Bold{\mu}}^{(\Bold{\lambda})}(z)
\end{equation}
where the CS models of the operators are now $N\times N$ matrices with $N=\mathrm{dim}_\C(\mathcal{W}_{(\Bold{\mu})})$. So, (\ref{second-order}) and its first-order equivalent contained in (\ref{group Heisenberg}) --- referred to an eigenstate basis in $\mathcal{H}$ --- looks like a (time-dependent) simple matrix model. This type of matrix equation is notoriously difficult to handle and yet simple enough that general qualitative information can be gleaned by inspection. Clearly $N$ can grow very large for multi-particle states, and for macro systems taking $N\rightarrow\infty$ is a reasonable approximation. So off hand, it appears that the $\mathfrak{Z}_+$ sector of SQM will look like a quantum membrane theory in $10$-d for multi-particle systems. Moreover, the VIPs of $ E_{ab}$ parametrize a subspace of the smooth manifold $Z$ --- so we know the starting point for the geometry of $\mathcal{X}\subset\sigma(\widehat{\mathfrak{e}}_{ab})$ if we know $(\psi_0(z)|\pi_z(J)\,\psi_0(z))_{\mathcal{W}_{z}}$.

For $\widetilde{\mathfrak{h}}_{\mathrm{U}}^\dag=\widetilde{\mathfrak{h}}_{\mathrm{U}}$ and suitable boundary conditions, (\ref{second-order}) can be formulated as a variation principle of the Lagrangian density (assuming time-independent $\widetilde{\mathfrak{h}}_{\mathrm{U}}$)
\begin{eqnarray}
\mathcal{L}_{\mathfrak{e}}(t)
&=&\frac{1}{2}\mathrm{Tr}
\left\{\left(\dot{\mathfrak{e}}_{ab}(t),\dot{\mathfrak{e}}^\dag_{cd}(t)\right)
+\left(ad(i\widetilde{\mathfrak{h}}_{\mathrm{U}})\mathfrak{e}_{ab}(t),
ad(i\widetilde{\mathfrak{h}}_{\mathrm{U}})^\dag\mathfrak{e}_{cd}^\dag(t)\right)\right\}\notag\\
&=:&\frac{1}{2}\mathrm{Tr}\left\{\left(D_t\mathfrak{e}_{ab}(t)\,
,D^\dag_t{\mathfrak{e}}^\dag_{cd}(t)\right)\right\}
\end{eqnarray}
where $\mathrm{Tr}(\cdot,\cdot)$ is a sesquilinear form on the Lie algebra, $D_t:=d/dt+ad(i\widetilde{\mathfrak{h}}_{\mathrm{U}})$, and $\widetilde{\mathfrak{h}}_{\mathrm{U}}^\dag=\widetilde{\mathfrak{h}}_{\mathrm{U}}$ was used in the second line. In particular, suppose $\widetilde{\mathfrak{h}}_{\mathrm{U}}\in\mathfrak{P}$ is pure `internal' gauge, i.e. $\widetilde{\mathfrak{h}}_{\mathrm{U}}\in\mathfrak{U}(4)$. Then $\widetilde{\mathfrak{h}}_{\mathrm{U}}\sim\sum_{a,b}\widetilde{\mathfrak{u}}_{ab}$ with $\widetilde{\mathfrak{h}}_{\mathrm{U}}^\dag=\widetilde{\mathfrak{h}}_{\mathrm{U}}$ as required, and it follows from $[\mathfrak{u}_{ab},\mathfrak{e}_{cd}]\in\mathfrak{Z}_+$ that $[d\mathfrak{e}_{ab}/dt,\mathfrak{e}_{ab}]=0$. At the other extreme, if the Hamiltonian is `external' in the sense that $\widetilde{\mathfrak{h}}_{\mathrm{U}}\sim\sum_{a,b}(\widetilde{\mathfrak{e}}^\dag_{ab}\pm\widetilde{\mathfrak{e}}_{ab})$, then $\left[[d\mathfrak{e}_{ab}/dt,\mathfrak{e}_{ab}],\mathfrak{e}_{ab}\right]=0$.

The Lagrangian density yields the evolution equation for $\mathfrak{e}^\dag_{ab}$ as well. Obviously, the CS model of the evolution equation of $E_{-a-b}$ referred to the eigenfunctions $\mathit{\Psi}^{(\Bold{\lambda})}$ is more complicated than simple matrix quantum mechanics. However, being mutually commuting, $E_{-a-b}$ possess a different eigenbasis where their evolution \emph{is} governed by simple matrix quantum mechanics --- modulo gauge symmetry considerations. The physical interpretation of such eigenstates is not clear: nevertheless, they may provide a possibly interesting dual picture.

The same Lagrangian density can be used for the entire Lie algebra:
\begin{proposition}\label{Lagrangian density}
Let $\mathfrak{g}_{ab}\in\mathfrak{Sp}(8)$ and suppose $\widetilde{\mathfrak{h}}_{\mathrm{U}}$ is time-independent. The Lagrangian density that generates evolution in $\mathfrak{Sp}(8)$ induced by $\widetilde{\mathfrak{h}}_{\mathrm{U}}$ is given by
\begin{equation}
\mathcal{L}_{\mathfrak{g}}(t)=\frac{1}{2}\mathrm{Tr}\left\{\left(D_t{\mathfrak{g}}_{ab}(t)\,
,D^\dag_t\mathfrak{g}^\dag_{cd}(t)\right)\right\}
\end{equation}
where $\mathrm{Tr}(\cdot,\cdot)$ is a sesquilinear form on $\mathfrak{Sp}(8)$ and $D_t:=d/dt+ad(i\widetilde{\mathfrak{h}}_{\mathrm{U}})$.
\end{proposition}
From here one can use the functional Mellin transform to realize, for example, the partition function associated with $\underline{\Bold{T}}$ (defined in the next subsection).

Alternatively, the Lagrangian density can be represented on the Hilbert space and expressed in terms of operators $G_{ab}:=\rho'(\mathfrak{g}_{ab})\in L(\mathcal{H})$ as
\begin{equation}\label{operator Lagrangian}
\mathcal{L}_{G}(t)=\frac{1}{2}\langle\psi_0|\left|D_t{G}_{ab}(t)\right|^2\,\psi_0\rangle\;.
\end{equation}
This suggests an analogous Lagrangian density formulation for the Heisenberg equation for bounded operators $O^{-1}_\lambda=\mathit{\Pi}^{-1}_\lambda(\mathrm{O})$ associated with observables $\mathrm{O}\in\mathbf{F}_{\mathbb{S}}(G^\C)$.
\begin{equation}
\mathcal{L}_{O^{-1}_\lambda}(t)=\frac{1}{2}\langle\psi_0| \left|D_t{O^{-1}_\lambda}(t)\right|^2\,\psi_0\rangle\;.
\end{equation}

Given that $Sp(8,\C)/U(4,\C)$ can be identified with the cotangent bundle $T^\ast Z$ for suitable $Z\subset Sp(8,\C)$, the form of these Lagrangians begs to formulate the dynamics of SQM as Hamiltonian mechanics of a ${U}(4)$ gauge theory on a non-commutative phase space. A proper and thorough treatment of this notion lies outside our present scope, and we will be content describing a coherent state phase space in the next section.

\subsection{CS phase space}
The definition of field was given in terms of CS parametrized by $10$ complex coordinates. We want to now express them in terms of $20$ real parameters.  For this purpose, use a matrix phase space CS model. This effectively separates the `internal' gauge freedom from the `external' gauge freedom and more closely corresponds to observed fields in terrestrial particle physics. Since we want a position/momentum interpretation, consider the coset space $Sp(8,\R)/U(4)$.

To construct the model, return to the complex setup and make use of remark \ref{phase space induction} to construct a phase space vector bundle $\mathcal{I}$ and its associated CS model by inducing a representation from $U(4)$. Define the operators
$Q_{ab}:=\rho'({\mathfrak{e}}_{ab}+{\mathfrak{e}}^\dag_{ab})/2$ to go along with
$\mathit{\Pi}_{ab}=\rho'(\mathfrak{e}_{ab}-\mathfrak{e}^\dag_{ab})/2$. Then a phase space CS can be defined by
 \begin{eqnarray}
|q,\pi\,;\widetilde{\Bold{\mu}})
&:=&\left(\exp\left\{\sum_{a,b}({z}_{ab}\mathfrak{e}_{ab}
+{z}_{ab}^\ast\mathfrak{e}^\dag_{ab})\right\}\right)
|\widetilde{\Bold{\mu}})\notag\\
&=&\left(\exp\left\{\sum_{a,b}q_{ab}({\mathfrak{e}}_{ab}+\mathfrak{e}^\dag_{ab})+
i\pi_{ab}(\mathfrak{e}_{ab}-\mathfrak{e}^\dag_{ab})\right\}\right)|\widetilde{\Bold{\mu}})
\end{eqnarray}
where ${z}_{ab}=q_{ab}+i\pi_{ab}$, $(q,\pi)\in Z\times Z^\ast$, and $\widetilde{\Bold{\mu}}\in \mathcal{W}_{(\Bold{\mu})}\oplus\mathcal{W}^\dag_{(\Bold{\mu})}$.
CS realizations of other phase space objects can be defined along the same lines as before. However, cross-sections of $\mathcal{I}$ cannot be directly identified with state vectors. For that we need to single out a distinguished polarization on $Sp(8,\C)/U(4,\C)$ that allows determination of a Lagrangian sub-space.\footnote{Explicitly, choose a connection on $Sp(8,\R)$. Use it to construct a completely integrable horizontal distribution $H\subset T(Sp(8,\R))$. Identify the horizontal sub-space with a suitable linear combination of ten mutually commuting generators in $\mathfrak{Z}_+\oplus\mathfrak{Z}_-$. Conclude that a connection is required to fix a polarization.} Then the coordinates on the sub-space can be consistently identified with the spectrum of ten mutually-commuting operators in $\mathfrak{Z}$. Consequently, given a polarization, the CS picture of a state vector is $\Bold{\psi}_{\widetilde{\Bold{\mu}}}(q,\pi)$, and the notion of elementary particles and fields applies also here.

As discussed previously, we want to associate generators of `external' dynamics with
\begin{equation}
\Bold{Q}+\Bold{\mathit{\Pi}}:=\left(
\begin{array}{llll}
Q_1 & Q_{12} & Q_{13} & Q_{14} \\
Q_{12} & Q_2 & Q_{23} & Q_{24} \\
Q_{13} & Q_{23} & Q_3 & Q_{34} \\
Q_{14} & Q_{24} & Q_{34} & Q_4 \\
                \end{array}
              \right)
+\left(\begin{array}{llll}
\mathit{\Pi}_1 & \mathit{\Pi}_{12} & \mathit{\Pi}_{13} &\mathit{\Pi}_{14} \\
\mathit{\Pi}_{21} & \mathit{\Pi}_2 & \mathit{\Pi}_{23} & \mathit{\Pi}_{24} \\
\mathit{\Pi}_{31} & \mathit{\Pi}_{32} & \mathit{\Pi}_3 & \mathit{\Pi}_{34} \\
\mathit{\Pi}_{41} & \mathit{\Pi}_{42} & \mathit{\Pi}_{43} & \mathit{\Pi}_4 \\
                \end{array}\right)\;,
\end{equation}
and generators of `internal' dynamics  with
\begin{equation}
\Bold{M}+\Bold{A}:=\left(
\begin{array}{llll}
M_1 & M_{12} & M_{13} & M_{14} \\
M_{12} & M_2 & M_{23} & M_{24} \\
M_{13} &M_{23} & M_3 & M_{34} \\
M_{14} & M_{24} & M_{34} & M_4 \\
\end{array}
\right)+\left(
\begin{array}{llll}
A_1 & A_{12} & A_{13} & A_{14} \\
A_{21} & A_2 & A_{23} & A_{24} \\
A_{31} &A_{32} & A_3 & A_{34} \\
A_{41} & A_{42} & A_{43} & A_4 \\
\end{array}
\right)
\end{equation}
where
\begin{eqnarray}
Q_{a}&=&\rho'(\mathfrak{e}_{a}+{\mathfrak{e}}_{a}^\dag)/2
\hspace{.4in}  \;\;\;\;\mathit{\Pi}_{a}=\rho'(\mathfrak{e}_{a}-{\mathfrak{e}}_{a}^\dag)/2\notag\\
Q_{a,b}&=&\rho'(\mathfrak{e}_{a,b}+{\mathfrak{e}}_{a,b}^\dag)/2
\hspace{.45in}\mathit{\Pi}_{a,b}=\rho'(\mathfrak{e}_{a,b}-{\mathfrak{e}}_{a,b}^\dag)/2
\end{eqnarray}
while the operators representing $\mathfrak{u}_{ab}$ are
\begin{eqnarray}
M_{a}&=&\rho'(\mathfrak{h}_{a}+\mathfrak{h}^\dag_{a})/2
\hspace{.8in}A_{a}
=\rho'(\mathfrak{h}_{a}-\mathfrak{h}^\dag_{a})/2=0\notag\\
M_{a,b}&=&\rho'(\mathfrak{e}_{a,-b}+\mathfrak{e}_{a,-b}^\dag)/2
\hspace{.45in} A_{a,b}=\rho'(\mathfrak{e}_{a,-b}-\mathfrak{e}_{a,-b}^\dag)/2\;.
\end{eqnarray}
 Note that
\begin{eqnarray}
\Bold{Q}^\dag
&=&\Bold{Q}\hspace{.5in}\Bold{\mathit{\Pi}}^\dag=-\Bold{\mathit{\Pi}}\notag\\
\Bold{M}^\dag &=&\Bold{M}\hspace{.45in}\Bold{A}^\dag=-\Bold{A}\;.
\end{eqnarray}

As expected, a gauge picture emerges with a choice of Lagrangian sub-space on the base manifold $(q,\pi)$. With the canonical choice, the CS models of $\{\mathit{\Pi}_{ab},M_{ab},A_{ab}\}$ become gauge potentials while the $Q_{ab}$ parametrize the configuration space. Interpret ${\underline{\mathit{\Pi}_{ab}}}:=\mathit{\Pi}_{ab}+{A}_{ab}$ as `interaction momentum' operators and
${\underline{Q_{ab}}}:=Q_{ab}+{M}_{ab}$
as `interaction configuration' operators. This motivates the definition
\begin{definition}\label{stress-energy}
The interaction stress-energy operator $\underline{\Bold{T}}\in L(\mathcal{H})$ is
defined by
\begin{equation}
{\underline{\Bold{T}}}:=\frac{1}{2}\left(\left\{\underline{\Bold{\mathit{\Pi}}},
\underline{\Bold{\mathit{\Pi}}}^\dag\right\}
+\left\{\underline{\Bold{Q}},\underline{\Bold{Q}}^\dag\right\}\right)\;.
\end{equation}
\end{definition}
In the CS matrix picture, $\widehat{\underline{\Bold{T}}}$ is a $4N\times4N$ second-order partial
differential operator where $N:=\mathrm{dim}_\C(\mathcal{W}_{(\widetilde{\Bold{\mu}})})$. For stress-energy CS eigenfunctions $\Bold{\psi}^{(\kappa)}_{\widetilde{\Bold{\mu}}}(q,\pi)$, this gives an interpretation of energy;
\begin{equation}
\mathrm{tr}\left(\widehat{\underline{\Bold{T}}}\right)\Bold{\psi}^{(\kappa)}_{\widetilde{\Bold{\mu}}}(q,\pi)
=:E_{\kappa}\,\Bold{\psi}^{(\kappa)}_{\widetilde{\Bold{\mu}}}(q,\pi)\;.
\end{equation}
In particular, the vacuum energy is, by definition,
\begin{equation}
\mathrm{tr}(\widehat{\underline{\Bold{T}}})\varphi_0=|\Bold{v}_\mu|^2\varphi_0=:E_0\,\varphi_0\;.
\end{equation}

A more informative object is the stress-energy CS eigenfunction  defined by
\begin{equation}
\widehat{\underline{\Bold{T}}}\Bold{\psi}^{(\Bold{\kappa})}_{\widetilde{\Bold{\mu}}}(q,\pi)
=\Bold{\kappa}\,\Bold{\psi}^{(\Bold{\kappa})}_{\widetilde{\Bold{\mu}}}(q,\pi)
\end{equation}
where $\Bold{\kappa}$ is a $4\times4$ matrix with real entries and $E_\kappa=\mathrm{tr}(\Bold{\kappa})$. Of course, evolution will generally alter the form of $\widehat{\underline{\Bold{T}}}$ rendering $\Bold{\psi}^{(\Bold{\kappa})}_{\widetilde{\Bold{\mu}}}(q,\pi)$ no longer stationary. But then one can define a new parabolic decomposition and retrace the quantization procedure to arrive at a new description of a quadratic Hamiltonian operator and its CS model of eigenstates. Needless to say, physical interpretation relative to the new parabolic decomposition may be highly nontrivial.

Like the symmetric, symplectic, and complex forms constructed from GIPs of pre-geometry; the GIP of stress-energy is actually a function of just $q$ due to gauge covariance of the ground state.  The form of $\mathfrak{Z}_+$ leads
naturally to the interpretation that $\underline{\Bold{T}}$, referred to a CS on the cotangent bundle diagonal \emph{near the ground state}, corresponds to a dynamically-induced, observed stress-energy
space-time \emph{tensor} $\underline{T_{ab}}(m)$ when $k\in\{1,3\}$. This is closely related to the symmetric, symplectic, and almost complex
structures defined on $\mathbb{M}^\C$ in subsection \ref{observed geometry}, and it immediately suggests
$\underline{\Bold{T}}$ as a potential generator of evolution.

\begin{remark}
Although this subsection dealt with phase space associated with real polarizations, similar considerations could be applied to holomorphic polarizations. Presumably, this would lead to a Segal-Bargmann holomorphic phase space construction. Of course these are only two of many possible choices of polarizations that may or may not be useful.
\end{remark}

\begin{remark}
Clearly the Standard Model symmetry group is contained in $U(4,\C)$, but if SQM is a realistic theory it must dictate exactly how $U(4,\C)$ reduces. Extracting those details is expected to be a major project far beyond our scope. Instead, we will be content to offer some observations.

A direct model starts with a Lagrangian subspace of $Sp(8,\R)/U(4)$  and assumes that the sixteen `internal' degrees of freedom associated with $U(4)$ of a quantum state $\breve{\psi}:\mathcal{P}\rightarrow\mathcal{W}_{(\mu)}$ are dynamically realized in the $(3,1)$ and $(2,2)$ regions of $\mathbb{M}[Sp(8,\R)]$. In these regions, only the unitary directions correspond to conserved charges and the sixteen charges of $U(4)$ become thirteen charges associated with $U(3)\times U(2)$ plus three leftover real parameters. These partially characterize the basic representations furnished by $\mathcal{W}_{(\mu)}$. (Recall the adjoint representation is a direct sum of the first three basic representations $\mathcal{V}_1^{(0)},\,\mathcal{V}_8^{(1)},\,\mathcal{V}_{27}^{(3)}$.) One then invokes spontaneous symmetry breaking to obtain $U(3)\times U(2)\rightarrow U(3)\times U(1)$. It is tempting to identify the $9+1$ group elements coming from $\mathfrak{U}(3)\oplus\mathfrak{U}(1)$ with the $SU(3)\times U(1)$ gauge bosons of the Standard model\footnote{There is, of course, an excess of one  gauge boson in the strong sector eventually requiring interpretation and/or explanation. We address this in a subsequent paper and argue that $U(3)$ is a viable alternative to $SU(3)$. Alternatively, of course, one could just assume $U(3)$ somehow is broken to $SU(3)$.} and the $16-(9+1)$ deficit of real degrees of freedom with the three complex massive electroweak boson states.

Alternatively, as $\mathfrak{U}(4,\C)$ contains two copies of $\mathfrak{U}(4)$, two possible subduction patterns are  $\mathfrak{U}(4)\rightarrow \mathfrak{U}(3)\oplus \mathfrak{U}(1)$ and $\mathfrak{U}(4)\rightarrow \mathfrak{U}(2)\oplus \mathfrak{U}(2)$. One might suppose that these patterns would be associated with overlapping regions in the  non-commutative phase space. An assumed Lagrangian slice  $X_D=Sp(8,\R)/(Sp(4,\R)\times U(4))$ of $Sp(8,\C)/U(4,\C)$ determined by the dynamics would presumably extrapolate between $U(3)\times U(1)$ and $U(2)\times U(2)$\footnote{ Evidently $\mathfrak{U}(2)\oplus \mathfrak{U}(2)$ would correspond to eight gauge bosons and four complex massive boson states. This contains some essence of technicolor.} --- settling on $U(3)\times U(1)$ in experimentally accessible regions of phase space. Note that $U(4)/(U(3)\times U(1))$ is isomorphic to projective twistor space.

\end{remark}

\subsection{Classical SQM}
Consider a closed quantum system in the CS phase space model with stress-energy $Ad(g)\underline{\Bold{T}}=:\underline{\Bold{T}}^{g}=:\underline{\Bold{T}}+\Bold{V}^{g}$. Assume CS eigenfunctions $\widehat{\underline{\Bold{T}}}\,\Bold{\psi}^{(\Bold{\kappa})}_{\widetilde{\Bold{\mu}}}(q,\pi)
=\Bold{\kappa}\Bold{\psi}^{(\Bold{\kappa})}_{\widetilde{\Bold{\mu}}}(q,\pi)$ that realize an over-complete set of eigenstates in $\mathcal{H}$ with a discrete spectrum (since the system is bounded). Because $\mathrm{tr}(\underline{\Bold{T}})\sim |\rho'(\mathfrak{h}_a)|^2$, these eigenfunctions are associated with elementary particles by definition.  Then, for a Schr\"{o}dinger CS sate-vector,
\begin{equation}
\Bold{\psi}_{\widetilde{\Bold{\mu}}}(q,\pi;t):=\sum_{\Bold{\kappa}} c_{\Bold{\kappa}}(t)\,\Bold{\psi}^{(\Bold{\kappa})}_{\widetilde{\Bold{\mu}}}(q,\pi)
=(q,\pi;\widetilde{\Bold{\mu}}|\,\sum_{\Bold{\kappa}} c_{\Bold{\kappa}}(t)\,\Bold{\psi}^{(\Bold{\kappa})}_{\widetilde{\Bold{\mu}}}\rangle\;.
\end{equation}
Use the CS resolution of the identity to write
\begin{equation}
c_{\Bold{\kappa}}(t)=\langle\Bold{\psi}^{(\Bold{\kappa})}_{\widetilde{\Bold{\mu}}}|\psi(t)\rangle
=\int_{U_i}{\Bold{\psi}^{(\Bold{\kappa})}_{\widetilde{\Bold{\mu}}}}^\dag(q,\pi)
\,\mathbf{d\,\Bold{\sigma}}(q,\pi)\,\Bold{\psi}_{\widetilde{\Bold{\mu}}}(q,\pi;t)\;.
\end{equation}
In particular,
\begin{eqnarray}
N_{\Bold{\kappa},\Bold{\kappa}}=N_{\Bold{\kappa},\Bold{\kappa}'}\,\delta(\Bold{\kappa},\Bold{\kappa}')
&:=&\int_{U_i}{\Bold{\psi}^{(\Bold{\kappa})}_{\widetilde{\Bold{\mu}}}}^\dag(q,\pi)
\,\mathbf{d\,\Bold{\sigma}}(q,\pi)\,{\Bold{\psi}^{(\Bold{\kappa}')}_{\widetilde{\Bold{\mu}}}}(q,\pi)\,\delta(\Bold{\kappa},\Bold{\kappa}')\notag\\
&=&\int_{U_i}{\Bold{\psi}^{(\Bold{\kappa})}_{\widetilde{\Bold{\mu}}}}^\dag(q,\pi)
\,\Bold{P}(q,\pi)\,{\Bold{\psi}^{(\Bold{\kappa}')}_{\widetilde{\Bold{\mu}}}}(q,\pi)\,\delta(\Bold{\kappa},\Bold{\kappa}')\;d(q,\pi)\notag\\
\end{eqnarray}
can be interpreted as the number density function in the spectrum
space of $\underline{\Bold{T}}$. Consequently, under suitable conditions ${\Bold{\psi}^{(\Bold{\kappa})}_{\widetilde{\Bold{\mu}}}}(q,\pi)$ can
be regarded as a CS distribution with an associated
probability measure
$\mu(q,\pi):={\Bold{\psi}^{(\Bold{\kappa})}_{\widetilde{\Bold{\mu}}}}^\dag(q,\pi)
\,\mathbf{\Bold{\sigma}}(q,\pi)\,{\Bold{\psi}^{(\Bold{\kappa})}_{\widetilde{\Bold{\mu}}}}(q,\pi)/N_{\Bold{\kappa},\Bold{\kappa}}$;
which means that $\mathrm{tr}({\Bold{\psi}^{(\Bold{\kappa})}_{\widetilde{\Bold{\mu}}}}^\dag(q,\pi)
\,\Bold{P}(q,\pi)\,{\Bold{\psi}^{(\Bold{\kappa})}_{\widetilde{\Bold{\mu}}}}(q,\pi))$ is the
number density of particle-type $\Bold{\kappa}$ on non-commutative phase space $(\Bold{Q},\Bold{\Pi})$.

Evidently $\Bold{\psi}_{\widetilde{\Bold{\mu}}}(q,\pi;t)$ is a time-dependent superposition of elementary particle state vectors so it permits the notion of particle creation and annihilation. To see this in more detail, it is convenient to go to the interaction picture defined
by
\begin{equation}
\Bold{\Psi}_{\widetilde{\Bold{\mu}}}(q,\pi;t)
:=e^{-i\,t\,\widehat{\underline{\Bold{T}}}}\,\Bold{\psi}_{\widetilde{\Bold{\mu}}}(q,\pi;t)
\end{equation}
and
\begin{equation}
\Bold{V}^{g}_I(t):=e^{-i\,t\,\underline{\Bold{T}}}\,\Bold{V}^{g}\,e^{\,i\,t\,\underline{\Bold{T}}}\;.
\end{equation}
Then
\begin{equation}\label{expansion eq.}
\p_{t}\,c_{\Bold{\kappa}}(t)
=\sum_{\Bold{\kappa}'}(\Bold{V}^{g}_I(t))_{\Bold{\kappa},\Bold{\kappa}'}
\,c_{\Bold{\kappa}'}(t)
\end{equation}
and
\begin{equation}
\frac{d\Bold{O}_I(t)}{dt}=i[\underline{\Bold{T}}\,,\Bold{O}_I(t)]
\end{equation}\label{interaction dynamics}
with $(\Bold{V}^{g}_I(t))_{\Bold{\kappa},\Bold{\kappa}'}
:=\langle\Bold{\psi}^{(\Bold{\kappa})}_{\widetilde{\Bold{\mu}}}|\Bold{V}^{g}_I(t)
|\Bold{\psi}^{(\Bold{\kappa}')}_{\widetilde{\Bold{\mu}}}\rangle$ together describe the dynamics.

There are two observations to make. First, recall the definition of $\Bold{P}(q,\pi)$: It is related to the CS reproducing kernel so it will be time-dependent in general. This follows because the adjoint group action that induces dynamics will \emph{potentially} induce a change in the reproducing kernel. Hence, the particle content of $\Bold{\psi}_{\widetilde{\Bold{\mu}}}(q,\pi;t)$ changes not only because the coefficients $c_{\Bold{\kappa}}(t)$ are time-dependent but also \emph{the number density} $\mathrm{tr}({\Bold{\psi}^{(\Bold{\kappa})}_{\widetilde{\Bold{\mu}}}}^\dag(q,\pi)
\,\Bold{P}(q,\pi)\,{\Bold{\psi}^{(\Bold{\kappa})}_{\widetilde{\Bold{\mu}}}}(q,\pi))$ \emph{can change}. A time-dependent number density on phase space is in stark contrast to elementary quantum mechanics, and it owes its existence to the non-abelian $U(4)$ subgroup of dynamical $Sp(8,\R)$ --- indeed, any CS eigenfunctions $\widehat{\underline{\Bold{T}}}\,\Bold{\psi}^{(\Bold{\kappa})}_{\widetilde{\Bold{\mu}}}(q,\pi)
=\Bold{\kappa}\Bold{\psi}^{(\Bold{\kappa})}_{\widetilde{\Bold{\mu}}}(q,\pi)$ that happen to be $U(4)$ singlet states have constant number density since the reproducing kernel is trivial in this case.

The second observation concerns the CS model of the interaction Heisenberg equation. In the eigenfunction basis, the operator is realized as a matrix that will become very large for  macroscopic systems. It is known that the commutator approaches the Poisson bracket in the limit as $N=\mathrm{dim}_\C(\mathcal{W}_{(\Bold{\mu})})\rightarrow \infty$. This brings us to the correspondence principle and classical mechanics (CM).

Posit the correspondence principle in the form $[\cdot,\cdot]\rightarrow\{\cdot,\cdot\}_{P.B}$ as $N\rightarrow\infty$. Reinstating Plank's constant this can be more meaningfully formulated as $\hbar/\mathrm{tr}(N_{\kappa,\kappa})\rightarrow0$. Then, passage from the quantum to a classical phase space description via the correspondence principle is
standard, but it yields a non-standard result in this context. To simplify notation
slightly, restrict to the case of $U(4)$ singlets and suppose the dynamics of a closed and bounded system
is determined by $\underline{\Bold{T}}^{g}=H(\Bold{Q},\Bold{\mathit{\Pi}})$;  a self-adjoint operator that is  bounded from below.
The semi-classical dynamics relative to a CS of particle-type $\Bold{\kappa}$ is
expressed by  matrix equations
\begin{equation}\label{matrix H}
\frac{d\Bold{\pi}^{(\Bold{\kappa})}(t)}{dt}
  =\{\Bold{\pi}(t),H(\Bold{q}(t),\Bold{\pi}(t))\}^{(\Bold{\kappa})}\;,\;\;\;
  \frac{d\Bold{q}^{(\Bold{\kappa})}(t)}{dt}
  =\{\Bold{q}(t),H(\Bold{q}(t),\Bold{\pi}(t))\}^{(\Bold{\kappa})}
\end{equation}
where
$\Bold{\pi}^{(\Bold{\Bold{\kappa}})}(t)
:=\left\langle{\Bold{\psi}^{(\Bold{\kappa})}_{\widetilde{\Bold{\mu}}}}|
\Bold{\mathit{\Pi}}(t)\,{\Bold{\psi}^{(\Bold{\kappa})}_{\widetilde{\Bold{\mu}}}}
\right\rangle/N_{\Bold{\kappa},\Bold{\kappa}}$\,and
$\Bold{q}^{(\Bold{\kappa})}(t):=\left\langle{\Bold{\psi}^{(\Bold{\kappa})}_{\widetilde{\Bold{\mu}}}}|
\Bold{Q}(t)\,{\Bold{\psi}^{(\Bold{\kappa})}_{\widetilde{\Bold{\mu}}}}
\right\rangle/N_{\Bold{\kappa},\Bold{\kappa}}$, and the semi-classical bracket relative
to the $\Bold{\kappa}$-type CS is defined by
\begin{eqnarray}
\{\Bold{a},\Bold{b}\}^{(\Bold{\kappa})}
&:=&\left\langle{\Bold{\psi}^{(\Bold{\kappa})}_{\widetilde{\Bold{\mu}}}}\left|
[\Bold{A},i\Bold{B}]\right.\,{\Bold{\psi}^{(\Bold{\kappa})}_{\widetilde{\Bold{\mu}}}}
\right\rangle/N_{\Bold{\kappa},\Bold{\kappa}}\notag\\
&=&\frac{1}{N_{\Bold{\kappa},\Bold{\kappa}}}
\int_{Z_\p}{\Bold{\psi}^{(\Bold{\kappa})}_{\widetilde{\Bold{\mu}}}}^\dag(q,\pi)\Bold{P}(q,\pi)
\left[\widehat{\Bold{A}}
,\widehat{i\Bold{B}}\right]{\Bold{\psi}^{(\Bold{\kappa})}_{\widetilde{\Bold{\mu}}}}(q,\pi)\,d(q,\pi)\;.\notag\\
\end{eqnarray}
This follows readily from
\begin{eqnarray}\label{force equation}
\frac{d\Bold{\pi}^{(\Bold{\kappa})}}{dt}
&:=&\left\langle{\Bold{\psi}^{(\Bold{\kappa})}_{\widetilde{\Bold{\mu}}}}\left|
\frac{d\Bold{\mathit{\Pi}}}{dt}\right.\,{\Bold{\psi}^{(\Bold{\kappa})}_{\widetilde{\Bold{\mu}}}}
\right\rangle/N_{\Bold{\kappa},\Bold{\kappa}}\notag\\
&=&\left\langle{\Bold{\psi}^{(\Bold{\kappa})}_{\widetilde{\Bold{\mu}}}}\left| \left[\Bold{\mathit{\Pi}},iH(\Bold{Q},\Bold{\mathit{\Pi}})\right]\right.\,{\Bold{\psi}^{(\Bold{\kappa})}_{\widetilde{\Bold{\mu}}}}
\right\rangle/N_{\Bold{\kappa},\Bold{\kappa}}\notag\\
&=&\frac{1}{N_{\Bold{\kappa},\Bold{\kappa}}}\int_{Z_\p}{\Bold{\psi}^{(\Bold{\kappa})}_{\widetilde{\Bold{\mu}}}}^\dag(q,\pi)
\,\Bold{P}(q,\pi)\left[\widehat{\Bold{\mathit{\Pi}}}
,\widehat{iH(\Bold{Q},\Bold{\mathit{\Pi}})}\right]
{\Bold{\psi}^{(\Bold{\kappa})}_{\widetilde{\Bold{\mu}}}}(q,\pi)\,d(q,\pi)\notag\\
&=:&\left\{\Bold{\pi},H(\Bold{q},\Bold{\pi})\right\}^{(\Bold{\kappa})}\;.
\end{eqnarray}
We interpret $H(\Bold{q},\Bold{\pi})$
as the classical (matrix) Hamiltonian for large $N_{\Bold{\kappa},\Bold{\kappa}}$.\footnote{Since the URs were built from $L^2(Z,\mathcal{W})$ equivariant maps, the
CS are properly normalized and the integral well defined (assuming the eigenfunctions ${\Bold{\psi}^{(\Bold{\kappa})}_{\widetilde{\Bold{\mu}}}}$ are indeed a complete set).} As $N\rightarrow\infty$ these reduce to classical dynamics with Poisson bracket $\{\cdot,\cdot\}_{P.B}$.

Evidently, for $N\rightarrow\infty$, the classical object $\Bold{\pi}^{(\Bold{\kappa})}(t)$ is a symmetric $4\times4$ matrix of momenta, and the classical equations
reduce to Hamilton's equations under special
conditions. For example, consider a macroscopic system with ground state $\psi_0$ and suppose  pre-geometry with $k\in\{1,3\}$. Assume the
evolution is `mild' in the sense that the system's wave function
${\Bold{\psi}^{(\Bold{\kappa})}_{\widetilde{\Bold{\mu}}}}(q,\pi)$ is annihilated by the off-diagonal
components of $\Bold{Q}$ and $\Bold{\mathit{\Pi}}$. Then (\ref{force equation}) reduces to
\begin{eqnarray}\label{Hamilton's eq.}
\frac{dp_a^{(\Bold{\kappa})}(t)}{dt}
&=&\{p_a(t),H_a(\Bold{q},\Bold{\pi})\}^{(\Bold{\kappa})}
=:F_a^{(\Bold{\kappa})}(q_a,p_a;t)\;,\;\;\;\;a\in\{1,2,3,4\}\notag\\\notag\\
\frac{dq_a^{(\Bold{\kappa})}(t)}{dt}
&=&\{q_a(t),H_a(\Bold{q},\Bold{\pi})\}^{(\Bold{\kappa})}\;.
\end{eqnarray}
Define the system
configuration space $\mathcal{Q}$  to be the diagonal
spectrum $\mathcal{Q}:= \sigma({Q}_{a})$.  Let $\mathcal{P}$ denote
the diagonal spectrum $\sigma({\mathit{\Pi}_{a}})$ and construct the associated observed
phase space\footnote{Note that the phase space will not be a
cotangent bundle in general.} $\mathcal{Q}\times\mathcal{P}$. If  observed phase space is smooth, equip $\mathcal{Q}$ with the metric $( g_{ab})_q$ now with Minkowski signature, and identify
evolution-time with proper time. Then (\ref{Hamilton's eq.}) can be interpreted as Hamilton's equations (relative to the bracket $\{\cdot,\cdot\}^{(\Bold{\kappa})}$) on the observed relativistic phase space $\mathcal{Q}\times\mathcal{P}$.\footnote{Remark that the semi-classical bracket is not necessarily the Poisson bracket --- although it shares identical algebraic properties and so must be proportional to it on $\mathcal{Q}\times\mathcal{P}$. Also, recall that $[\mathfrak{q}_a,\mathfrak{p}_b]=\delta_{ab}\mathfrak{h}_a$ so $\{q_a,p_b\}^{(\Bold{\kappa})}=\delta_{ab}\|\mathfrak{h}_a\|^{(\Bold{\kappa})}$. This justifies calling $\mathcal{Q}\times\mathcal{P}$ the classical phase space according to the correspondence principle.}

With these notions of momentum and force defined on
$\mathcal{Q}\times\mathcal{P}$, the Boltzmann equation for a CS
probability distribution obtains as the classical reduction of the
Heisenberg equation for the density operator $\Bold{\rho}$. Explicitly,
\begin{equation}\label{Vlasov}
\frac{d f^{(\Bold{\kappa})}(\Bold{q},\Bold{\pi};t)}{dt}
:=\mathrm{tr}\,\frac{d\Bold{\rho}^{(\Bold{\kappa})}(t)} {dt}
=\mathrm{tr} \left\{\Bold{\rho},{H}({\underline{\Bold{q}}},
\underline{\Bold{\pi}})\right\}^{(\Bold{\kappa})}
\end{equation}
where now
${H}={H}({\underline{\Bold{q}}},
\underline{\Bold{\pi}})$
is a function of the `interaction position' and `interaction momentum' which includes the $U(4)$ contribution to $\underline{\Bold{T}}$ and presumably
encodes the classical body forces. However, more interesting than this scalar function equation is the $4\times 4$ matrix Boltzmann equation
\begin{equation}
\frac{d\Bold{\rho}^{(\Bold{\kappa})}(t)} {dt}
=\left\{\Bold{\rho},{H}({\underline{\Bold{q}}},
\underline{\Bold{\pi}})\right\}^{(\Bold{\kappa})}\;.
\end{equation}
We do not pursue it here, but off-hand it appears to include \emph{independent} linear and rotational classical degrees of freedom. In particular, for quasi particles of a meso/macroscopic systems, perhaps the matrix Boltzmann equation provides a handle on vortex dynamics.

\section{Conclusion}

There are three main pillars supporting SQM; dynamical $Sp\,(8,\R)$ to model quantum kinematics and
govern evolution, a coherent state arena for observation and
interpretation, and a ground state with memory that records external
interactions. We expect $SO(9,\R)$ to describe some
physical systems, but the implications were not pursued here.\footnote{It is
satisfying that $Sp\,(8,\C)$ and $SO(9,\C)$ are Langlands duals, so one might hope that
$SO\,(9,\C)$ describes strongly coupled SQM and vice versa.} We also suspect $Sp\,(2n,\R)$ and $SO(2n+1,\R)$ for $n<4$ may be
relevant symmetries for some ordered meso/macroscopic systems (i.e. ponderable media).\footnote{There is obviously plenty of  motivation to also consider $OSp(9|8)$, and one can see some remarkable similarities (or perhaps just coincidences) with certain conjectured M-theory constructs\cite{BFFS} in this case.}

Defining the symplectic quantum theory more or less follows standard quantum
mechanics except that the group determines both observables
associated with internal degrees of freedom and the kinematic
observables usually associated with phase space --- and the
commutation relations among them. The theory is quantized by
constructing the Hilbert space from induced representations and
using the functional Mellin transform to transfer the algebra of
observables to the operator algebra on the Hilbert space.

An obvious parabolic decomposition of $\mathfrak{Sp}(8,\R)=\mathfrak{Z}_+\oplus \mathfrak{P}$
hints at the possibility of a nontrivial internal /external symmetry
unification. According to our interpretation
of the algebra decomposition, space-time intervals are ground state
expectation values of certain quantum observables and Poincar\'{e} is a
rather special limiting symmetry. In consequence, meshing Poincar\'{e} with gauge quantum mechanics is not an issue, and the no-go theorems regarding mixing internel/Lorentz symmetries do not apply. In a nutshell, our proposal is to replace relativistic quantum mechanics with symplectic quantum mechanics; eventually including $SO(9,\R)$ and perhaps $OSp(9|8)$ as a portal between the two.

If the idea that $Sp(8,\R)$ and $SO(9,\R)$ are dual via a running coupling parameter is correct, the matrix gauge QM that includes them both could be described by a Lagrangian density that looks very much like a supersymmetric matrix model. However, the hypothesized underlying \emph{dynamical} group would be $Sp(8,\R)\times SO(9,\R)$. So, although there would necessarily be some kind of matching conditions for either description at some fixed value of the coupling parameter, there would be no need to introduce superpartners. Nevertheless, the Lagrangian would appear to posses a BRST-like symmetry.\cite[\S4]{LA4} Our hunch is that this BRST-like symmetry and the physical meaning of the conjectured duality can be (partially) understood \emph{at the semi-classical level} with the help of skew-Gaussian and Liouville functional integrals developed in \cite[\S\S4,5]{LA4}. The semi-classical implication is: Weakly coupled $Sp(8,\R)$ describes \emph{scattering} while weakly coupled $SO(9)$ describes \emph{correlations}. They are two sides of the same coin, so to speak. From the $Sp(8,\R)$ side, weakly coupled systems can be viewed as separate states evolving over time while strongly coupled systems only manifest time-dependent correlations --- in the latter case there is no sense of time-evolution relative to some scattering configuration/phase space.\footnote{At the quantum level, there is always time-evolution of operators. But according to our interpretation, for strongly coupled systems it is not always manifested as CS dynamics at the semi-classical level. This highlights an observational disconnect between  evolution-time at the quantum level and  evolution-time at the semi-classical level. Perhaps there is something to learn about reversible/irreversible `time' from this.} The BRST-like symmetry enjoyed by the Lagrangian formulation that includes them both exposes the (hypothesized) syplectic/orthogonal nature of quantum systems and justifies the dual description.\footnote{One wonders if this might be the essence of particle/wave duality.}

It is significant that the dynamics of $\mathfrak{Z}_+\oplus \mathfrak{P}$ is
governed by the same group they help generate. In fact, for any dynamical evolution of a system, $\mathfrak{P}$ will in general change according to the adjoint group action. From this
perspective, it is natural to guess that the ten generators
contained in $\mathfrak{Z}_{-}$ have something to do with inertia. Similarly, the ten generators comprising $\mathfrak{Z}_+$ generate a $10$-d configuration space that we interpret as four linear dimensions and six directed-area `dimensions'.\footnote{There are clearly implications regarding entropy content and transfer if a volume element in a $4$-d space can be characterized by the six directed-area `dimensions' --- especially if their associated observables encode vortex-like dynamics.} These momentum-type and position-type operators will not remain static in general. It is tempting to interpret this as quantum
gravity --- at least as a toy model.

The coherent states facilitate physical interpretation, so it might be enlightening to develop the matrix CS model of the dynamics by quantizing a $P^\C$ gauge theory on the base space $\Bold{Z}$ to see if it matches SQM. A functional integral approach seems to be indicated: but it properly deserves a detailed study and so was not included in this paper.\footnote{Our $C^\ast$-algeba with $\ast$-convolution is a far cry from the algebra of matrix valued functions with the Moyal product so, off hand, the standard approach\cite{WS,DN} does not seem to apply here.} Of course such a bottom-up approach might not agree with the original quantum theory. If it did agree, it would likely improve physical understanding. Presumably, the construction would look similar to a QFT-like gauge theory on $\Bold{Z}$ but with important differences; the most obvious being non-local observables, no \emph{a priori} Poincar\'{e} symmetry, and an adjustable, evolution-dependent vacuum.

There is a growing consensus in recent years that space-time and gravity are emergent in some sense and matrix models are perhaps more fundamental than Yang-Mills QFT (for a review see \cite{ST}). It is remarkable that the fairly simple-minded SQM leads naturally and unambiguously (modulo the initial choice of dynamical group)  to similar notions. It is likely that quantum mechanics based on the sister group $SO(9,\R)$ and parent group $OSp(9|8)$ have more surprises in store.

\section{Acknowledgment}
I am grateful to Jean-Pierre Zablit for collaborative discussions over a five year period beginning in 2006. What began as an
attempt to model certain nonlinear optical systems ultimately hinted that $Sp(8,\R)$ and $SO(9,\R)$ were somehow fundamental.

\appendix
\section{Matrix CS representation}
This appendix derives the matrix CS model of the operators associated with $\mathfrak{Sp}(8)$ and the reproducing kernel on $Z$.(c.f. \cite{DQ})

By definition, a matrix CS is $\psi_{\Bold{\mu}}(\Bold{Z})=(\Bold{Z};\Bold{\mu}|\psi\rangle=(\Bold{\mu}|
e^{\frac{1}{2}\mathrm{tr}(\Bold{Z}\mathfrak{E}_-)}\,\psi\rangle\in U_i\times\mathcal{W}_{(\Bold{\mu})}$ where $(\Bold{Z})_{ab}:=(1+\delta_{ab})z_{ab}=:Z_{ab}$ and $\mathfrak{E}_-$ is a $4\times4$ symmetric operator-valued matrix with entries $\mathfrak{e}^\dag_{ab}$ with $a\leq b\in\{1,2,3,4\}$. This implies
\begin{eqnarray}
\widehat{\mathfrak{e}}^\dag_{ab}\,\psi_{\Bold{\mu}}(\Bold{Z})
&=&(\Bold{Z};\Bold{\mu}|{\mathfrak{e}}^\dag_{ab}\,\psi\rangle\notag\\
&=&(\Bold{\mu}|e^{\frac{1}{2}\mathrm{tr}(\Bold{Z}\mathfrak{E}_-)}{\mathfrak{e}}^\dag_{ab}\,
e^{-\frac{1}{2}\mathrm{tr}(\Bold{Z}\mathfrak{E}_-)}
e^{\frac{1}{2}\mathrm{tr}(\Bold{Z}\mathfrak{E}_-)}\,\psi\rangle\notag\\
&=&(\Bold{\mu}|\{Ad(e^{\frac{1}{2}\mathrm{tr}(\Bold{Z}\mathfrak{E}_-)}){\mathfrak{e}}^\dag_{ab}\}\,
e^{\frac{1}{2}\mathrm{tr}(\Bold{Z}\mathfrak{E}_-)}\,\psi\rangle\notag\\
&=&(\Bold{\mu}|{\mathfrak{e}}^\dag_{ab}\,
e^{\frac{1}{2}\mathrm{tr}(\Bold{Z}\mathfrak{E}_-)}\,\psi\rangle\notag\\
&=&(\frac{\p}{\p z_{ab}}\otimes I_{(\Bold\mu)})\,\psi_{\Bold{\mu}}(\Bold{Z})
\end{eqnarray}
where $ I_{(\Bold\mu)})$ is the identity in $\mathcal{W}_{(\Bold{\mu})}$. In the fourth line we used (\ref{commutation relations}) and the BCH formula
\begin{equation}
 Ad(e^{\frac{1}{2}\mathrm{tr}(\Bold{Z}\mathfrak{E}_-)}){\mathfrak{e}}^\dag_{ab}
 =\sum_{n=0}^\infty\frac{[(\frac{1}{2}\mathrm{tr}(\Bold{Z}\mathfrak{E}_-))^n,
 {\mathfrak{e}}^\dag_{ab}]}{n!}
 =\sum_{n=0}^\infty\sum^4_{c,d=1}
 \frac{1}{2^{n}n!}[(Z_{cd}\mathfrak{e}^\dag_{dc})^n,{\mathfrak{e}}^\dag_{ab}]\;.
\end{equation}
This can be written as
\begin{equation}
 \widehat{\mathfrak{E}}_{-}\,\psi_{\Bold{\mu}}(\Bold{Z})
 =\left(\p_{\Bold{Z}}\otimes I_{(\Bold\mu)}\right)\,\psi_{\Bold{\mu}}(\Bold{Z})
\end{equation}
where
\begin{equation}
 \p_{\Bold{Z}}:=\left(
\begin{array}{cccc}
\frac{\p}{\p z_{1}} & \frac{\p}{\p z_{12}} &  \frac{\p}{\p z_{13}} & \frac{\p}{\p z_{14}}\\ \\
\frac{\p}{\p z_{12}} & \frac{\p}{\p z_{2}} & \frac{\p}{\p z_{23}} &\frac{\p}{\p z_{24}}\\ \\
 \frac{\p}{\p z_{13}} & \frac{\p}{\p z_{23}} & \frac{\p}{\p z_{3}} & \frac{\p}{\p z_{34}} \\ \\
 \frac{\p}{\p z_{14}} & \frac{\p}{\p z_{24}} &\frac{\p}{\p z_{34}} & \frac{\p}{\p z_{4}} \\
  \end{array}
  \right)\;.
\end{equation}
Similarly,
\begin{equation}
 Ad(e^{\frac{1}{2}\mathrm{tr}(\Bold{Z}\mathfrak{E}_-)}){\mathfrak{u}}_{ab}
=\mathfrak{u}_{ab}+1/2\{\sum_cZ_{ca}\mathfrak{e}^\dag_{bc}+\sum_dZ_{da}\mathfrak{e}^\dag_{bd}\}
\end{equation}
yields
\begin{eqnarray}
\widehat{\mathfrak{u}}_{ab}\,\psi_{\Bold{\mu}}(\Bold{Z})
&=&(\Bold{\mu}|\{{\mathfrak{u}}_{ab}+(\Bold{Z}{\mathfrak{E}}_-)_{ab}\}\,
e^{\frac{1}{2}\mathrm{tr}(\Bold{Z}\mathfrak{E}_-)}\,\psi\rangle\notag\\
&=&\sum_{\Bold{\mu}'}
\left(((\Bold{U})_{ab})_{\Bold{\mu}\Bold{\mu}'}+(\Bold{Z}\p_{\Bold{Z}})_{ab} \delta_{\Bold{\mu}\Bold{\mu}'}\right)\,\psi_{\Bold{\mu}'}(\Bold{Z})
\end{eqnarray}
where $(\Bold{U})_{ab}:=\bar{\varrho}'(\mathfrak{u}_{ab})$ is a known UIR of $U(4)$. The matrix form is
\begin{equation}
 \widehat{\mathfrak{E}}_{U}\,\psi_{\Bold{\mu}}(\Bold{Z})
 =\left(\Bold{Z}\p_{\Bold{Z}}\otimes I_{(\Bold{\mu})}+\mathcal{U}\right)\cdot\,\psi_{\Bold{\mu}'}(\Bold{Z})
\end{equation}
with $(\mathcal{U})_{ab,\Bold{\mu}\Bold{\mu}'}:=((\Bold{U})_{ab})_{\Bold{\mu}\Bold{\mu}'}$.
Finally,
\begin{equation}
 Ad(e^{\mathrm{tr}(\Bold{Z}\mathfrak{E}_-)}){\mathfrak{e}}_{ab}
={\mathfrak{e}}_{ab}-1/2\{(\Bold{Z}\Bold{U})_{ab}+(\Bold{Z}\Bold{U})_{ba}\}
+1/2\{(\Bold{Z}\Bold{Z}{\mathfrak{E}}_-)_{ab}
+(\Bold{Z}\Bold{Z}{\mathfrak{E}}_-)_{ba}\}
\end{equation}
gives
\begin{eqnarray}
 \widehat{\mathfrak{E}}_{+}\,\psi_{\Bold{\mu}}(\Bold{Z})
 &=&\left((\Bold{Z}\Bold{Z}\p_{\Bold{Z}})_{sym}\otimes I_{(\Bold{\mu})}-(\Bold{Z}\mathcal{U})_{sym}\right)\cdot\,\psi_{\Bold{\mu}'}(\Bold{Z})\notag\\
 &=&\left((\Bold{Z}\p_{\Bold{Z}}-5\Bold{I})\Bold{Z}\otimes I_{(\Bold{\mu})}-(\Bold{Z}\mathcal{U})_{sym}\right)\cdot\,\psi_{\Bold{\mu}'}(\Bold{Z})
\end{eqnarray}
where $\Bold{Z}\mathcal{U}=(\Bold{Z}\Bold{U})_{\Bold{\mu}\Bold{\mu}'}$ and we used $[(\p_{\Bold{Z}})_{ab},(\Bold{Z})_{cd}]=\delta_{ac}\delta_{bd}+\delta_{ad}\delta_{bc}$ to write
\begin{equation}
\sum_c(\Bold{Z})_{ac}(\p_{\Bold{Z}})_{cb}=\sum_c(\p_{\Bold{Z}})_{bc}(\Bold{Z})_{ca}-5\delta_{ab}\;.
\end{equation}
Observe that $\widehat{\mathfrak{E}}_{-}$ and $\widehat{\mathfrak{E}}_{+}$ are both symmetric matrices with respect to $U_i\in Z$.

Now we find the normal-ordered form of the reproducing kernel on $M_4^{sym}(\C)\times\mathcal{W}_{(\Bold{\mu})}$
\begin{eqnarray}\label{explicit overlap}
(\Bold{K}(\Bold{Z}',\Bold{Z}^\ast))_{\Bold{\mu}'\,\Bold{\mu}}
&:=&({\Bold{Z}'};\Bold{\mu}'|\Bold{Z}^\ast;\Bold{\mu})
=(\Bold{\mu}'|e^{\frac{1}{2}\mathrm{tr}(\Bold{Z}'\mathfrak{E}_-)}
\,e^{\frac{1}{2}\mathrm{tr}(\Bold{Z}^\ast\mathfrak{E}_+)}|\Bold{\mu})\;.
\end{eqnarray}
As $Sp(8,\C)$ is a simple Lie group, we can proceed in any faithful representation. It is convenient to work in the defining representation  where $(\Bold{\mathfrak{E}}_+)_{ab}
=\mathrm{Id}_{4+a,b}+\mathrm{Id}_{4+b,a}$, and $(\Bold{\mathfrak{E}}_-)_{ab}
=-(\Bold{\mathfrak{E}}_+)_{ab}$, and $(\Bold{\mathfrak{E}}_U)_{ab}
=\mathrm{Id}_{2+b,2+a}-\mathrm{Id}_{a,b}$, and the $4\times4$ matrix $\mathrm{Id}_{a,b}$ has $1$ at position $(a,b)$ and $0$ everywhere else. In this representation, $\Bold{\mathfrak{E}}_+$ and $\Bold{\mathfrak{E}}_-$ are strictly triangular matrices and hence
\begin{equation}
e^{\frac{1}{2}\mathrm{tr}(\Bold{Z}'\mathfrak{E}_-)}
=\left(
   \begin{array}{cc}
     \Bold{I} & \Bold{0} \\
     \Bold{Z}'& \Bold{I} \\
   \end{array}
 \right)\;,\;\;\;\;\;\;\;\;
e^{\frac{1}{2}\mathrm{tr}(\Bold{Z}^\ast\mathfrak{E}_+)}
=\left(
   \begin{array}{cc}
     \Bold{I} & -\Bold{Z}^\ast \\
     \Bold{0}& \Bold{I} \\
   \end{array}\right)\;.
\end{equation}
Since  $\Bold{\mathfrak{E}}_U$ is $4\times4$ block diagonal, for some $(\Bold{B})_{ab}=B_{ab}$,
\begin{equation}
\mathrm{tr}(\Bold{B}\Bold{\mathfrak{E}}_U)
=\left(
   \begin{array}{cc}
   \Bold{B} & \Bold{0} \\
     \Bold{0} & -\Bold{B}^{\mathrm{T}} \\
   \end{array}
 \right)
\end{equation}
and
\begin{equation}
e^{\mathrm{tr}(\Bold{B}\Bold{\mathfrak{E}}_U)}
=\left(
   \begin{array}{cc}
     e^{\Bold{B}} & \Bold{0} \\
     \Bold{0} & e^{-\Bold{B}^{\mathrm{T}}} \\
   \end{array}\right)\;.
\end{equation}
Hence, from the BCH expansion we have
\begin{equation}
\Bold{K}(\Bold{Z}',\Bold{Z}^\ast)
=\left(
   \begin{array}{cc}
     \Bold{I} & -\Bold{Z}^\ast \\
     \Bold{Z}' & \Bold{I}-\Bold{Z}'\Bold{Z}^\ast \\
   \end{array}
 \right)
 =\left(e^{\frac{1}{2}\mathrm{tr}(\Bold{A}\Bold{\mathfrak{E}}_+)}
 \;e^{\mathrm{tr}(\Bold{B}\Bold{\mathfrak{E}}_U)}
 \;e^{\frac{1}{2}\mathrm{tr}(\Bold{C}\Bold{\mathfrak{E}}_-)}\right)
\end{equation}
for some $4\times4$ matrices $\Bold{A},\Bold{B},\Bold{C}$. Performing the matrix multiplication on the right-hand side and equating $4\times4$ blocks yields
\begin{eqnarray}
\Bold{I}&=&e^{\Bold{B}}-\Bold{A}\,e^{-\Bold{B}^{\mathrm{T}}}\,\Bold{C}\notag\\
\Bold{Z}^\ast&=&\Bold{A}\,e^{-\Bold{B}^{\mathrm{T}}}\notag\\
\Bold{Z}'&=&e^{-\Bold{B}^{\mathrm{T}}}\,\Bold{C}\notag\\
\Bold{1}-\Bold{Z}'\Bold{Z}^\ast&=&e^{-\Bold{B}^{\mathrm{T}}}\;.
\end{eqnarray}
Together, these conditions imply $\Bold{B}^{\mathrm{T}}=\log(\Bold{I}-\Bold{Z}'\Bold{Z}^\ast)^{-1}$, $\Bold{A}=\Bold{Z}(\Bold{I}-\Bold{Z}'\Bold{Z}^\ast)^{-1}$, and $\Bold{C}=(\Bold{I}-\Bold{Z}'\Bold{Z}^\ast)^{-1}\Bold{Z}'$ as long as $\Bold{Z}'\Bold{Z}^\ast\in B^{open}_1(\Bold{I})$. Conclude that for any representation $\rho^{(r)}$,
\begin{equation}
(\Bold{K}^{(r)}(\Bold{Z}',\Bold{Z}^\ast))_{\Bold{\mu}'\Bold{\mu}}
=(\Bold{\mu}'|
e^{\mathrm{tr}(\Bold{B}(\Bold{Z}',\Bold{Z}^\ast)
\,{\rho^{(r)}}'(\Bold{\mathfrak{E}}_U))}
|\Bold{\mu})\;.
\end{equation}

\section{Emergent Einstein-Yang-Mills}
We have seen the semi-classical limit of SQM leads to a dynamical configuration space-time (under suitable conditions). But this doesn't necessarily mean the dynamics is governed by general relativity. This appendix aims to sketch an argument with \emph{no claim to proof} (with some observations and remarks along the way) that SQM contains the Einstein-Yang-Mills action. Indeed, suitable expectation values of the generators of $Sp(8,\R)$ contain the necessary ingredients, and constructing the relevant action is almost trivial. The interpretation is that Einstein-Yang-Mills is the `expected dynamics' of a particular Hamiltonian operator with respect to a certain class of coherent states. The choice of Hamiltonian and coherent states is not dictated by SQM but we will see that they are natural in a sense (from a semi-classical viewpoint).

The bulk of the argument lies in showing our quantization procedure\cite{LA1} is well defined. Since our approach differs from generally accepted quantization methods, this step is not trivial. The main hurdle is justifying the functional Mellin transform and this is presented already in \cite{LA2}.  It turns out that since the $C^\ast$-algebra is non-commutative and the underlying group is non-abelian, functional Mellin provides a representation precisely when $\alpha=1$. We will show that, when $\alpha=1$, functional Mellin is essentially a crossed product. Crossed products are rigorously established so functional Mellin is on solid ground for $\alpha=1$. Hence, in place of functional Mellin, the quantum theory can be just as well formulated in the crossed product formalism --- which is certainly well defined.

But let's first get to Einstein-Yang-Mills. To begin, note the potential energy term in the matrix quantum mechanics encoded in (\ref{operator Lagrangian}) is minimized for the case of diagonal evolution operators acting on diagonal $G_{ij}$. It is natural then to restrict attention to states $\psi_{\mathrm{diag}}$ represented by CS with $\mathrm{supp}(\psi_{\Bold{\mu}}(\Bold{Z}))\subseteq\sigma(Q_a)$ and $M_a$ eigenvalues associated with $\mathcal{G}(1,3)$ which, assuming sufficient topological and smoothness conditions, allows to interpret $\sigma(Q_a)$ as coordinates on a complex space-time manifold $\M^\C$. Construct the principal bundle $(\sigma(\mathcal{P}),\sigma(Q_a),\breve{pr},P^\C)$ where $\sigma(\mathcal{P})$ is the spectra of $\rho(G^\C)$ (also assumed to be a sufficiently smooth topological space).

We have already seen that $\langle J\rangle_{\mathrm{diag}}$ and $\langle g_{ab}\rangle_{\mathrm{diag}}$ allow to construct (for suitable $\sigma(Q_a)$) a real $4$-d manifold $\M$ with metric $g_{ab}(m)$. Since $\langle g_{ab}\rangle_{\mathrm{diag}}=\langle \{\mathit{\Pi}_{a},{\mathit{\Pi}_{b}^\dag}\}\rangle_{\mathrm{diag}}$, one can interpret $\Bold{e}^a:=\langle \mathit{\Pi}_{a}\rangle_{\mathrm{diag}}\in T^\ast\sigma(\mathcal{P})$  as a tetrad which induces a linear connection $\nabla$ with connection coefficients $\Bold{\gamma}$ that allow to identify $T_m\M$ with $\mathrm{span}_\R\{\mathfrak{e}^\dag_a-\mathfrak{e}_a\}$. Choosing the compatibility condition $\nabla \tilde{\Bold{e}}^a=0$ (where $\tilde{\Bold{e}}^a:=s_i^\ast{\Bold{e}}^a$ is the pullback of $\Bold{e}^a$ to $\M$ by a local section $s_i^\ast$) partially fixes the gauge leaving six momentum-type gauge symmetries. Since the commutator represents a derivation on the Lie algebra  and the expectations $\langle\mathit{\Pi}_{a,b}\rangle_{\mathrm{diag}}\in T^\ast\sigma(\mathcal{P})$ pull back to Lie algebra-valued $1$-forms on $\M$, the obvious candidate for a spin connection is $\Bold{\omega}^{ab}:=\langle \mathit{\Pi}_{a,b}\rangle_{\mathrm{diag}}$ with curvature form $\Bold{R}^{ab}(\Bold{\omega})= {d}\langle \mathit{\Pi}_{a,b}\rangle_{\mathrm{diag}}
+\langle[\mathit{\Pi}_{a,b},{\mathit{\Pi}_{a,b}^\dag}]\rangle_{\mathrm{diag}}$. There remain two momentum-type expectations to characterize --- $\langle[\mathit{\Pi}_{a},{\mathit{\Pi}^\dag_{b}}]\rangle_{\mathrm{diag}}$ and $\langle \{\mathit{\Pi}_{a,b},{\mathit{\Pi}_{a,b}^\dag}\}\rangle_{\mathrm{diag}}$. The first defines the antisymmetric product of tetrads $\Bold{e}^{a}\barwedge \Bold{e}^{b}:=\langle [\mathit{\Pi}_{a},{\mathit{\Pi}^\dag_{b}}]\rangle_{\mathrm{diag}}$. The second vanishes on the class of CS currently under consideration but contributes to the metric for generic CS supported on the entire spectrum $\sigma(Q)$.

The remaining non-momentum-type expectations $\Bold{u}^{ab}:=\langle M_{ab}+A_{ab}\rangle_{\mathrm{diag}}=:\langle U_{ab}\rangle_{\mathrm{diag}}$ can be interpreted as $U(4)$ gauge connections whose field strength $\Bold{F}^{ab}(\Bold{u})$ is associated with `internal' degrees of freedom. The anti-symmetric form $\langle[U_{ab},U_{ab}^\dag]\rangle_{\mathrm{diag}}$ together with the exterior derivative of $\Bold{u}$ determine $\Bold{F}^{ab}(\Bold{u})$ while the symmetric combination $\langle\{U_{ab},U_{ab}^\dag\}\rangle_{\mathrm{diag}}$ contributes to the stress-energy.  The term `internal' is supported by the fact that the connection and the complex structure $\langle J\rangle_{\mathrm{diag}}$ afford an orthogonal decomposition of the real subalgebra $\Re(\mathfrak{P}^\C)=T_m\M\oplus\mathrm{span}_\R\{{\mathfrak{e}_{a,b}^\dag}-\mathfrak{e}_{a,b}\}
\oplus\mathrm{span}_\R\{\mathfrak{u}_{ab}\}$.

Accordingly, for CS supported only on the diagonal of $M_4^{sym}(\C)$, expectations of $\rho'(\mathfrak{P})$ generate geometric objects on $\sigma(Q_a)$ associated with dynamical curvature and internal symmetry that is generated by a \emph{mixture} of $\mathit{\Pi}_{a,b}$ and $U_{ab}$. The action functional using these ingredients that yields gravity and Yang-Mills dynamics is already well-known;
\begin{equation}
S(\Bold{e},\Bold{\omega},\Bold{u})=\int_{\sigma(Q_a)}\mathfrak{tr}\left\{(\Bold{e}\barwedge \Bold{e}) \wedge \Bold{R}(\Bold{\omega})
+\Lambda ((\Bold{e}\barwedge \Bold{e})\wedge (\Bold{e}\barwedge \Bold{e}))+\Bold{F}(\Bold{u})\wedge\ast\Bold{F}(\Bold{u})\right\}
\end{equation}
where $\mathfrak{tr}\{\cdot\}$ denotes the Killing form on $\mathfrak{G}$ and $g_{ab}(m)$ is used to define the wedge product $\wedge$ and Hodge star $\ast$.

Recall the variation of $S(\Bold{e},\Bold{\omega},\Bold{u})$ implies vanishing torsion which in turn restricts $\Bold{\omega}$ to be the Levi-Civita connection. So, although the internal quantum symmetry is a mixture of $\mathit{\Pi}_{a,b}$ and $U_{ab}$, the semi-classical symmetry settles on $\mathfrak{so}(3,1)$ plus whatever the Yang-Mills e.o.m. implies. In this regard, an equally valid\footnote{There are certainly other consistent gauge choices beyond the two presented here.} starting point is to partially fix the gauge by choosing $\langle\mathit{\Pi}_{a,b}\rangle_{\mathrm{diag}}$ to generate internal(local) Lorentz symmetry. Then the remaining momentum-type degrees of freedom $\langle \mathit{\Pi}_{a}\rangle_{\mathrm{diag}}$ (which in the CS picture are responsible for translation on $\M$) will lead to teleparallel gravity where the role of curvature and torsion are reversed.

We have not yet derived the notion of mass from SQM, but insofar as this notion characterizes the interaction of matter with $\langle\mathit{\Pi}_{ab}\rangle_{\mathrm{diag}}$, these constructions reveal that the equivalence principle is a semi-classical manifestation of the gauge symmetry generated by $\mathit{\Pi}_{ab}$. In other words, SQM implies the equivalence principle --- assuming our argument can be strengthened to proof and  the notion of mass can be extracted. (This hints maybe mass is a semi-classical notion and the distinction of inertial and gravitational mass is an artifact of gauge choice.)

The standard arguments that lead to this (more or less canonical) action functional are just as valid here, and in this sense Einstein-Yang-Mills emerges naturally from SQM.\footnote{One hopes that, under closer inspection, the quantum dynamics inevitably lead to this action functional augmented with matter field terms at low energies.} But just to be thorough, one should imbed $\sigma(Q_a)\hookrightarrow Z$ and then work backwards using (\ref{operator symbol}) to construct the operator $H(\mathit{\Pi},M,A)$ such that $\langle H(\mathit{\Pi},M,A)\rangle_{\mathrm{diag}}=S(\Bold{e},\Bold{\omega},\Bold{u})$. Then gauge invariance at the quantum level requires $\langle \delta H(\mathit{\Pi},M,A)\rangle_{\mathrm{diag}}=0$ which yields the appropriate classical field equations for $\Bold{R}(\Bold{\omega})$ and $\Bold{F}(\Bold{u})$. But remind that this holds only for the special class of states $\psi_{\mathrm{diag}}$ for which $\sigma(Q_a)$ has the manifold structure of space-time.

Repeating this exercise for general states; under suitable conditions we arrive at a real $10$-d manifold $\sigma(Q)\hookrightarrow M_4^{sym}(\C)$ with metric $\Bold{g}_{ab}(\Bold{Z}):=s^\ast_i\langle\{\mathit{\Pi}_{ab},\mathit{\Pi}_{ab}^\dag\}\rangle$ and inner product $(\Bold{Z}^1|\Bold{Z}^2):=\mathrm{tr}(\Bold{Z}^1\Bold{g}\Bold{Z}^2)$ on $TM_4^{sym}(\C)$. Use the metric to endow $\sigma(Q)$ with the Riemannian connection and scalar curvature $\Bold{R}$. There is a decad of $1$-forms $\Bold{e}^{ab}:=\langle\mathit{\Pi}_{ab}\rangle$; but instead\footnote{One could consider using $\Bold{e}^{ab}$ as vielbein like in the $4$-d case. The compatibility condition $\Bold{\nabla}\Bold{e}^{ab}=0$ would completely fix the gauge leaving no remaining `internal' momentum-type gauge symmetries. Of course both approaches are valid, but each leads to a different description of ``reality'' at the semi-classical level.} of using it to solder $T_Z\sigma(Q)$ to a subspace of $\mathfrak{P}^\C$, we consider it a gauge connection with field strength $\Bold{P}^{ab}(\Bold{e})
={d}\langle\mathit{\Pi}_{ab}\rangle+\langle[\mathit{\Pi}_{ab},\mathit{\Pi}^\dag_{ab}]\rangle$. For the internal $U(4)$ symmetry we still have $\Bold{u}^{ab}=\langle U_{ab}\rangle$ gauge connections with field strength $\Bold{F}^{ab}(\Bold{u})$. Observe the external symmetry $\mathrm{span}_\R\{{\mathfrak{e}_{a,b}^\dag}-\mathfrak{e}_{a,b}\}$ of the $4$-d case gets augmented to $\mathrm{span}_\R\{{\mathfrak{e}_{ab}^\dag}-\mathfrak{e}_{ab}\}$ in $10$-d. The action functional of interest is
\begin{equation}
S(\Bold{g},\Bold{e},\Bold{u})=\int_{\sigma(Q)}\mathfrak{tr}\left\{\Bold{R}
+\Bold{P}^2(\Bold{e})+\Bold{F}^2(\Bold{u})+\Bold{T}\right\}\,d\Bold{\tau}
\end{equation}
where $\Bold{T}$ comes from the stress-energy (\ref{stress-energy}) and  $d\Bold{\tau}$ is the invariant measure induced by $\Bold{g}$. This action functional deserves a separate detailed investigation.

Turn now to the proposed method of quantization. Remind that the rationale for breaking with convention is to construct a quantum theory without first assuming a classical configuration/phase space. The conviction that functional integrals (which experience has shown capture quantum physics) should be formulated on topological groups along with the prominence of symmetry in QM lead naturally to the functional Mellin quantization scheme introduced in \cite{LA1}.

But there is another $C^\ast$-algebra approach based on crossed products that is on sound mathematical footing. The ingredients necessary to define crossed products\cite{W} are: i) a ``dynamical system'' $(A,G,\varepsilon)$ where $A$ is a $C^\ast$-algebra, $G$ is a locally compact group, and $\varepsilon:G\rightarrow Aut(A)$ is a continuous homomorphism; ii) some Hilbert space $\mathcal{H}$; iii) a representation $\varpi:A\rightarrow L_B(\mathcal{H})$; and iv) a unitary representation $U:G\rightarrow U(\mathcal{H})$. The two representations are required to satisfy the `covariance condition'
\begin{equation}\label{covariance condition}
\varpi(\varepsilon_g(a))=U_g\varpi(a)U_g^\ast\;,\;\;\;\;g\in G\;,\;\;a\in A\;.
\end{equation}
With these objects, a $\ast$-representation on $\mathcal{H}$ of $C_c(G,A)$ (continuous compact morphisms $\mathrm{f}:G\rightarrow A$) is supplied by the integral
\begin{equation}\label{crossed product}
\varpi\rtimes U(\mathrm{f}):=\int_G\varpi(\mathrm{f}(g))U_g\;d\mu(g)
\end{equation}
where $\mathrm{f}\in C_c(G,A)$ and $\mu$ is a Haar measure on $G$.

A product and involution are introduced on $C_c(G,A)$ according to
\begin{equation}
(\mathrm{f}_1\ast \mathrm{f}_2)(g):=\int_G \mathrm{f}_1(\tilde{g})\varepsilon_{\tilde{g}}(\mathrm{f}_2(\tilde{g}^{-1}g))\;d\mu(\tilde{g})
\end{equation}
and
\begin{equation}
\mathrm{f}^\ast(g):=\Delta(g^{-1})\varepsilon_g(\mathrm{f}(g^{-1})^\ast)
\end{equation}
where $\Delta$ is the modular function on $G$. Completion of $C_c(G,A)$ with respect to the norm defined by
\begin{equation}
\|\mathrm{f}\|:=\mathrm{sup}\|\varpi\rtimes U(\mathrm{f})\|
\end{equation}
is a $C^\ast$-algebra called the crossed product denoted by $A\rtimes_\varepsilon G$.

The crucial property of this construction is a one-to-one correspondence between non-degenerate covariant representations associated with $(\varpi,U)$ and non-degenerate representations of  $A\rtimes_\varepsilon G$ which preserve direct sums, irreducibility, and equivalence. So the $C^\ast$-algebra $A\rtimes_\varepsilon G$ can be used to model the $C^\ast$-algebra encoded in the dynamical system $(A,G,\varepsilon)$ endowed with a covariant representation $(\varpi,U)$. We recognize the covariant condition as an algebra automorphism by a group element; which, in particular, for the evolution operator in quantum mechanics becomes the integrated Heisenberg equation.

In practice, crossed products are invariably (as far as we are aware) applied to a classical dynamical system $(C_0(X),G,\varepsilon)$ where $C_0(X)$ is the commutative algebra of complex valued continuous functions vanishing at infinity and $X$ is some Hausdorff topological space with a $G$-action encoded in $\varepsilon$. But the construction of crossed products allows other possibilities. Suppose $A$ is non-commutative and $G$ is a locally compact (sub)group of units. Then $G$ acts on $A$ by inner automorphisms which means the covariance condition is automatic and $\varepsilon$ is unneeded. Setting $\varepsilon\equiv Id$ brings the involution and product of crossed products into agreement with functional Mellin. Then insisting that $\mathrm{f}$ be equivariant saves the $C^\ast$-algebra structure of $A\rtimes_{Id} G$. In this situation then, $\mathbf{F}_{\mathbb{S}}(G_\lambda^\C)\cong A\rtimes_{Id} G$ and representations furnished by functional Mellin are in one-to-one relation to $A\rtimes_{Id} G$ and therefore in one-to-one relation to the (non-classical) dynamical system $(A,G,Id)$. Accordingly, the quantum kinematics and dynamics of functional Mellin and crossed products are equivalent in this case.

This indicates functional Mellin can be made mathematically rigorous. To render the construction complete, however, requires one more step to connect to physics; the idea(assumption) that \emph{observables} coming from $\mathbf{F}_{\mathbb{S}}(G_\lambda^\C)$ are actually ``projections'' of elements in $\mathbf{F}_{\mathbb{S}}(G^\C)$. That is, the quantum system is encoded in $\mathbf{F}_{\mathbb{S}}(G^\C)$ and the Born rule is a natural consequence of the existence of surjective homomorphisms from $G^\C$ to locally compact $G_\lambda^\C$.\footnote{If one restricts to unitary representations, a probability interpretation can be attached to the measures associated with $G_\lambda^\C$.} Granted this assumption, we have a legitimate quantum theory (see \cite{LA1} for details).


\begin{thebibliography}{99}
\bibitem{LA1}
J. LaChapelle, Functional Integral Approach to $C^*$-algebraic Quantum Mechanics I: Heisenberg and Poincar\'{e}, arXiv:math-ph/1505.08102 (2015).


\bibitem{LA4}
J. LaChapelle, Functional Integration on Topological Groups I: Non-Gaussian Functional Integrals with Applications,  arXiv:math-ph/1501.01602 (2015).

\bibitem{PR}
P. Pajas and R. Raczka, Degenerate Representations of the Symplectic
Groups I. the Compact Group $Sp(n)$, \emph{International Center for
Theoretical Physics}, Trieste (1966).

\bibitem{RW}
D.J. Rowe and J.L. Wood, \emph{Fundamentals of Nuclear Models:
Foundational Models}, World Scientific, New Jersey, (2010).

\bibitem{ADMS}
B. Arvind, N. Dutta, and R. Simon, The Real Symplectic Groups in Quantum Mechanics and Optics, arXiv:quant-ph/9509002 (1995).

\bibitem{WU}
A. W\"{u}nsche, Symplectic groups in quantum optics, \emph{J. Opt. B} \textbf{2}, (2000).

\bibitem{KM}
N. Krausz and M.S. Marinov, Exact evolution operator on non-compact
group manifolds, \emph{J. Math. Phys.} \textbf{41}, 5180 (2000).

\bibitem{C-B}
Y. Choquet-Bruhat and C. DeWitt-Morette, \emph{Analysis, Manifold
and Physics}, Elsevier, Amsterdam, (1982).

\bibitem{FU}
J. Fuchs, and C. Schweigert, \emph{Symmetries, Lie Algebras and
Representations}, Cambridge Univ. Press, Cambridge UK, (2003).

\bibitem{JQ}
Jin-Quan Chen, Jialun Ping,  and Fan Wang, \emph{Group
Representation Theory for Physicists}, World Scientific, New Jersey,
(2002).

\bibitem{CM}
S. Coleman and J. Mandula, All Possible Symmetries of the S Matrix,
\emph{Phys. Rev.} \textbf{159}(5), 1251 (1967).

\bibitem{DQ}
J. Deenen and C. Quesne, Partially coherent states of the real
symplectic group, \emph{J. Math. Phys.} \textbf{25}(8), 2354 (1984).

\bibitem{BR}
S.D. Bartlett, D.J. Rowe, and J. Repka, Vector coherent state
representations, induced representations, and geometric
quantization: II. Vector coherent state representations,
arXiv:quant-ph/0201130v2.


\bibitem{W}
D.P. Williams,  \emph{Crossed Products of $C^\ast$-algebras}.
American Mathematical Society, Providence, Rhode Island, (2007).


\bibitem{LA2}
J. LaChapelle, Functional Integration on Topological Groups II: Functional Mellin Transforms, arXiv:math-ph/1501.01889 (2015).

\bibitem{L} N.P. Landsman, Rieffel induction as generalized quantum
Marsden-Weinstein reduction, \emph{J. Geo. and Phys.} 15, 285-319,
(1995).


\bibitem{M1}
G.W. Mackey, On induced representations of groups, \emph{Amer. J. Math.} \textbf{73}, 576--592, (1951).

\bibitem{M2}
G.W. Mackey, Induced representations of locally compact groups. I., \emph{Ann. of Math.} \textbf{55}(2) 101--139, (1952).

\bibitem{M3}
G.W. Mackey, Induced representations of locally compact groups. II. The Frobenius reciprocity theorem, \emph{Ann. of Math.} \textbf{58}(2) 193--221, (1953).

\bibitem{KT}
E. Kaniuth, K.F. Taylor, Induced Representations of Locally Compact Groups, Cambridge Univ. Press, Cambridge, UK, (2013).

\bibitem{KW}
R.C. King and B.G. Wybourne, Holomorphic discrete series and
harmonic series unitary irreducible representations of non-compact
Lie groups: $Sp\,(2n,\R)$, $U(p,q)$, and $SO^\ast(2n)$, \emph{J.
Phys. A: Math. Gen.} \textbf{18}, 3113 (1985).

\bibitem{GG}
I.M. Gelfand  and M.I. Graev , \emph{Am. Math. Soc. Transl.} Ser. 2
\textbf{164}, 116--46 (1967).

\bibitem{HC}
Harish-Chandra, Discrete series for semisimple Lie groups: I,
\emph{Acta Math.} \textbf{113}, 241--318 (1965),and Discrete series
for semisimple Lie groups: II, \emph{Acta Math.} \textbf{116}, 1-111
(1966).
\bibitem{HS}
H. Hecht and W. Schmid, A proof of Blattner's conjecture,
\emph{Inventiones Math.} \textbf{31}, 129--154 (1975).

\bibitem{AS}
M. Atiyah and W. Schmid, A geometric construction for the discrete
series for semisimple Lie groups, \emph{Inventiones Math.}
\textbf{42}, 1--62 (1977).

\bibitem{KP}
P. Kramer and Z. Papadopolos, Hilbert spaces of analytic functions
and representations of the positive discrete series of $Sp\,(6,\R)$,
\emph{J. Phys. A: Math. Gen.}  \textbf{19}, 1083--1092 (1986).

\bibitem{HN}
J. Hilgert and K.-H. Neeb, Structure and Geometry of Lie Groups, Springer, New York (2012).

\bibitem{CA}
R. Camporesi, Harmonic Analysis and Propagators on Homogeneous
Spaces, \emph{Phys. Rep.} \textbf{196}, 1--134 (1990).


\bibitem{AN}
V.I. Arnol'd and A.B. Givental', Symplectic Geometry, in \emph{Dynamical Systems IV}, Springer-Verlag, New York (1990).

\bibitem{WN}
J. Wei and E. Norman, On global representations of the solutions of linear differential equations as a product of exponentials, \emph{Proc. Am. Math. Soc.} \textbf{4}(4), 575--581 (1963).


\bibitem{G}
R. Gilmore, \emph{Lie Groups, Lie Algebras, and Some of Their Applications}. Dover Publications, New York (2002).

\bibitem{LA3}
J. LaChapelle, Functional Integration on Constrained Function
Spaces I and II, arXiv:math-ph/1212.0502 (2012) and arXiv:math-ph/1405.0461 (2014).


\bibitem{CA/D-W3}
P. Cartier  and C. DeWitt-Morette, \emph{Functional Integration:
Action and Symmetries}. Cambridge University Press, Cambridge (2006).

\bibitem{BL}M. Blau and G. Thompson, Localization and
Diagonalization, \emph{J. Math. Phys.} \textbf{36}, 2192 (1995).

\bibitem{BFFS}
T. Banks, W. Fischler, S.H. Shenker, and L. Susskind, M Theory as a Matrix Model: a Conjecture, \emph{Phys. Rev. D} \textbf{55}, 5112 (1997).

\bibitem{ST}
H. Steinacker, Emergent geometry and gravity from matrix models: an introduction, HAL Id: hal-00606646 (2011).

\bibitem{WS}
N. Seiberg and E. Witten, String theory and noncommutative geometry, \emph{JHEP}\textbf{09(1999)32} (1999).

\bibitem{DN}
M.R. Douglas and N.A. Nekrasov, Noncommutative Field Theory, \emph{Rev. Mod. Phys.} \textbf{73}, 977 (2001).


\end{thebibliography}
\end{document}